\documentstyle[preprint,aps]{revtex}
\tighten
\begin{document}
\newcommand{\rb}[1]{\raisebox{-1ex}[-1ex]{#1}}
\newcommand{\sst}[1]{\scriptscriptstyle{#1}}
\newcommand{\dfrac}[2]{\mbox{$\displaystyle\frac{#1}{#2}$}}

\draft

\title{
The nonmesonic weak decay of the hypertriton
}

\author{J.~Golak$^\dagger$, K.~Miyagawa$^\ddagger$,
        H.~Kamada$^*$, H.~Wita\l a$^\dagger$,
        W.~Gl\"ockle}
\address{Institut f\"ur Theoretische Physik II,
         Ruhr Universit\"at Bochum, 44780 Bochum, Germany}
\address{$^\dagger$ Institut of Physics, Jagellonian University,
                    PL 30059 Cracow, Poland}
\address{$^\ddagger$ Department of Applied Physics,
                     Okayama University of Science,
                     Ridai-cho Okayama 700, Japan}
\address{$^*$ Paul Scherrer Institute,
              CH-5232 Villigen PSI, Switzerland}

\author{A.~Parre\~no$^{**}$, A.~Ramos$^{**}$,
        C.~Bennhold$^{***}$}
\address{$^{**}$ Departament d'Estructura i Constituents de la Mat\`eria,
Facultat de F\'{\i}sica, Diagonal 647, E-08028, Spain} 
\address{$^{***}$ Center for Nuclear Studies,
                  The George Washington University, Washington D.C., USA}

%\date{}
\maketitle

\begin{abstract}

The nonmesonic decay of the hypertriton is calculated based on a hypertriton
wavefunction and 3N scattering states, which are rigorous solutions of
3-body Faddeev equations using realistic NN and hyperon-nucleon interactions.
The pion-exchange together with heavier meson exchanges
for the $\Lambda N \rightarrow N N $ transition is considered.
The total nonmesonic decay rate is found to be 0.5\% of the free
$\Lambda$ decay rate.  Integrated
as well as differential decay rates are given. The p-- and n-- induced decays
are discussed thoroughly and it is shown that the corresponding total
rates cannot be measured individually.
\end{abstract}

\narrowtext

\section{Introduction}
\label{secIN}

The hypertriton consisting of a neutron, proton and a $\Lambda$ or $\Sigma$,
which strongly convert into each other, is bound against $\Lambda$-deuteron
decay by 0.13 $\pm$ 0.05 MeV. Recently that number could be
reproduced~\cite{1} by solving the Faddeev equations based on realistic
NN forces and the Nijmegen hyperon-nucleon interaction~\cite{2}.
The resulting wavefunction has all two-body correlations exactly built in
as enforced by the various two-body forces.   As the lightest hypernucleus 
the hypertriton plays the same role in hypernuclear physics that the
deuteron does 
in nuclear physics.  However,  in contrast to the deuteron, the hypertriton
is subject to the weak decay and has a lifetime comparable to that of the free
$\Lambda$, 
$\tau_\Lambda=2.63 \times 10^{-10}$ seconds.
The first data on light hypernuclear
lifetimes have been obtained using bubble chamber experiments and emulsion
works which in most cases only detected the mesonic decay modes.
These measurements suffered from low precision, very poor statistics
and difficulties with particle identification, leading to fairly large error
bars. 
Along with the mesonic two-body decay mode
$^{3}_{\Lambda }H \rightarrow \pi^- (\pi^0) + ^3He (^3H)$
there are the corresponding mesonic multi-body decay modes
$^{3}_{\Lambda }H \rightarrow \pi^- (\pi^0) + d + p(n)$ and
$^{3}_{\Lambda }H \rightarrow \pi^- (\pi^0) + p + n + p(n)$. 
The most precise experiment to date for the combined two- and multi-body decay
modes gave a value of
$\tau=(2.28+0.46-0.33) \times 10^{-10}$ seconds~\cite{3}.
A similar measurement furthermore was also able to estimate the decay branching ratio 
$\Gamma (^{3}_{\Lambda }H \rightarrow \pi^-  + ^3He) /
 \Gamma(^{3}_{\Lambda }H \rightarrow$
all  $\pi^-$ mesonic modes) as $0.30 \pm 0.07$.

Besides the mesonic decay channels there are also two nonmesonic modes,
$^{3}_{\Lambda }H \rightarrow  d + p$ and
$^{3}_{\Lambda }H \rightarrow  p + p + n$.
While it is well known that these channels
dominate the weak decay of heavy hypernuclei they
are expected to be rare for the hypertriton since the
mesonic modes are not Pauli suppressed~\cite{4}.
As a first step this two-baryon decay mode $\Lambda N \to NN$ can be
understood in terms of the free-space
decay mechanism $\Lambda \rightarrow \pi N$ with virtual pion that is absorbed
on a second nucleon bound in the hypernucleus~\cite{5}.
 However, the large momentum
transfer involved in the reaction leads to a mechanism that is sensitive to
the short distance behavior of the amplitude and allows for
the exchange of heavier mesons. The
production of these mesons would be below threshold for the free-space
$\Lambda$ decay, but they can contribute through virtual exchange in a
two-baryon decay channel.

The weak nonmesonic decay channel is important since
it allows access to the fundamental aspects of the four-fermion, strangeness
changing weak interaction.
In general, starting with the Standard Model
electroweak Hamiltonian and taking into account
QCD corrections at short distances yields an effective weak
$V-A$ interaction that could presumably
predict the relative strength of the $\Delta S$=0 and $\Delta S$=1
transition. Thus, hadronic weak matrix elements of the form
$\langle M B' | H_w \mid B  \rangle$ can be calculated~\cite{6}.
Using these weak vertices as a starting point for effective
nuclear two-body operators that are then implanted into the nucleus
with the usual nuclear many-body wave functions
provides the testing ground for the effective interaction.

Parity violation in hadronic systems represents a unique tool
to study aspects of the nonleptonic weak interaction between hadrons.
The nonmesonic process
resembles the weak $\Delta$S=0 nucleon-nucleon interaction that has been
explored experimentally in parity-violating NN scattering
measurements.
However, the $\Lambda N \to NN$ two-body decay mode
can explore both the parity-conserving (PC) and the parity-violating
(PV) sector of the $\Delta$S=1 weak baryon-baryon interaction while in
the weak NN system one is limited to the weak PV interaction.
A number of theoretical approaches to the $\Lambda N \to NN$ decay mode have
been developed over the last thirty years which are more extensively
reviewed in Ref.~\cite{5}.
The $\Delta S=0$ weak
nucleon-nucleon interaction at low and intermediate energies has generally
been described in a meson exchange model involving one strong interaction
vertex and one weak one; the same approach can be used for a
microscopic description of the $\Delta S$=1 $\Lambda N \to NN$ mechanism.

A recently completed major study of the nonmesonic
decay of p-shell hypernuclei~\cite{7}
found that proper short-range correlations in the initial and final
state are of major importance in predicting decay rates and asymmetry
observables.  However, in a shell-model framework bound state wave
functions, spectroscopic factors, short-range correlations and final state
interactions do not all originate from the same underlying dynamics and,
therefore, introduce approximations that may be difficult to quantify.
Since the aim of investigating the nonmesonic decay is to extract
information on hadronic weak vertices
from the $\Lambda N \rightarrow NN$ process
the decay of few-body hypernuclei offers a window since all nuclear
structure ingredients are derived from the same baryon-baryon interaction.

It thus appears worthwhile to repeat a former study~\cite{4} on the weak
nonmesonic decay of the hypertriton, where a simplified uncorrelated
deuteron-$\Lambda$ wavefunction has been used. We expect that
correlations should play an important role, since the mesons emitted by the
weak hyperon-nucleon transition are reabsorbed by the nucleons.
The resulting meson-exchange operator acts like a two-body force
and consequently probes the hypertriton wavefunction in its dependence
on the pair distance between a hyperon and a nucleon.
Furthermore, the final three nucleons will interact strongly with each
other, which might influence significantly the decay process.
This dynamical ingredient has been neglected in Ref.~\cite{4}
and will now be fully incorporated.

In section~\ref{secII} the theoretical
formalism for the evaluation of the decay matrix element will be given.
Section~\ref{secIII} describes the necessary technicalities.
A special section~\ref{secIV} is devoted to the meson-exchange operator.
We present our results in section~\ref{secV}. We summarize and conclude
in section~\ref{secVI}.

\section{Formalism}
\label{secII}

There are two nonmesonic decay channels
\[
  {}^3_{\Lambda}H \ \longrightarrow \ \left\{ \
   \begin{array}{c}
   {\rm n + d} \\[4pt]
   {\rm n + n + p}
   \end{array} \right.
\]
According to standard rules, the partial decay probabilities in the total
momentum zero frame are
\[
d \, \Gamma^{\rm n+d} \ = \ \frac12 \, \sum_{m \, m_N \, m_d} \
\mid < \Psi^{(-)}_{ {\vec k}_N \, {\vec k}_d \,  m_N \, m_d }
\mid {\hat O} \mid \Psi_{ {}^3_\Lambda H \, m } >  \mid^2 \,
\]
\begin{equation}
d {\vec k}_N \, d  {\vec k}_d \, 2 \pi \,
\delta ( {\vec k}_N +  {\vec k}_d ) \,
\delta \left( M_{ {}^3_\Lambda H } - M_N - M_d - \,
\frac { {\vec k}_N^{\ 2} }{2 M_N} - \,
\frac { {\vec k}_d^{\ 2} }{2 M_d} \right)
\label{e1}
\end{equation}
and
\[
d \, \Gamma^{\rm n+n+p} \ = \ \frac12 \, \sum_{m \, m_1 \, m_2 \, m_3} \
\mid
< \Psi^{(-)}_{ {\vec k}_1 \, {\vec k}_2 \, {\vec k}_3 \, m_1 \, m_2 \, m_3 }
\mid {\hat O} \mid \Psi_{ {}^3_\Lambda H \, m } >  \mid^2 \,
\]
\begin{equation}
d {\vec k}_1 \, d {\vec k}_2 \, d {\vec k}_3 \, 2 \pi \,
\delta ( {\vec k}_1 + {\vec k}_2 + {\vec k}_3 ) \,
\delta \left( M_{ {}^3_\Lambda H } - 3 M_N \, -
\frac { {\vec k}_1^{\ 2} }{2 M_N} - \,
\frac { {\vec k}_2^{\ 2} }{2 M_N} - \,
\frac { {\vec k}_3^{\ 2} }{2 M_N} \right){\rm ,}
\label{e2}
\end{equation}
where $\Psi^{(-)}$ are appropriate three-nucleon scattering states,
$ {\hat O} $ the transition operator and $\Psi_{ {}^3_\Lambda H}$
the hypertriton wavefunction. We use throughout nonrelativistic
notation. The binding energies are defined as usual in terms
of various masses as
\begin{eqnarray}
 & M_{ {}^3_\Lambda H} \, = \, 2 M_N + M_\Lambda + \epsilon \nonumber \\[4pt]
 & M_d \, = \, 2 M_N + \epsilon_d
\label{e3}
\end{eqnarray}
Further we introduce Jacobi momenta for the final 3N states
\begin{eqnarray}
 & {\vec p} \, = \, {1 \over 2} \ ({\vec k}_1 - {\vec k}_2) \nonumber \\[4pt]
 & {\vec q} \, = \, {2 \over 3} \ ({\vec k}_3 -
{1 \over 2} \ ({\vec k}_1 + {\vec k}_2))
\label{e4}
\end{eqnarray}
and identify for the nd breakup ${\vec k}_3 = {\vec k}_N$
and ${\vec k}_1 + {\vec k}_2 = {\vec k}_d $.
Then some simple algebra leads to
\begin{equation}
d \,\Gamma^{\rm n+d} \ = \ \frac12 \, \sum_{m \, m_N \, m_d} \
\mid < \Psi^{(-)}_{ {\vec q}_0  \, m_N \, m_d }
\mid {\hat O} \mid \Psi_{ {}^3_\Lambda H \, m } >  \mid^2 \,
2 \pi \, \frac{2 M_N}{3} q_0 d {\hat q} {\rm ,}
\label{e5}
\end{equation}
with
\begin{equation}
q_0 \, = \, \sqrt{  \frac{4 M_N}{3} ( M_\Lambda - M_N + \epsilon - \epsilon_d)}
{\rm .}
\label{e6}
\end{equation}
Because of the averaging over spin directions the matrix element
squared is independent of ${\hat q}_0$ and we get just a number
\begin{equation}
\Gamma^{\rm n+d} \ = \ 8 \, \pi^2 \,  \frac{2 M_N}{3} \, q_0 \,
\frac12 \, \sum_{m \, m_N \, m_d} \
\mid < \Psi^{(-)}_{ {\vec q}_0  \, m_N \, m_d }
\mid {\hat O} \mid \Psi_{ {}^3_\Lambda H \, m } >  \mid^2 {\rm .}
\label{e6.5}
\end{equation}
Similar steps lead to
\begin{equation}
d \, \Gamma^{\rm n+n+p} \ = \ \frac12 \, \sum_{m \, m_1 \, m_2 \, m_3} \
\mid
< \Psi^{(-)}_{ {\vec p} \, {\vec q} \, m_1 \, m_2 \, m_3 }
\mid {\hat O} \mid \Psi_{ {}^3_\Lambda H \, m } >  \mid^2 \,
2 \pi \, \frac{2 M_N}{3}
\, q d {\hat q}
\, d {\hat p} p^2 d p
\label{e7}
\end{equation}
with
\begin{equation}
q \, = \, \sqrt{  \frac{4 M_N}{3} \left( M_\Lambda - M_N + \epsilon
- \frac { {\vec p}^{\ 2} }{M_N} \right)}
\label{e8}
\end{equation}
Again the spin-averaged part depends only on the angle $\theta$
between $\hat p$ and $\hat q$, thus
\begin{equation}
d \, \Gamma^{\rm n+n+p} \ = \
16 \, \pi^3 \, \frac23 \, M_N \,
q \, p^2 \, d p \, \sin \theta \, d \theta \,
\frac12 \, \sum_{m \, m_1 \, m_2 \, m_3} \
\mid
< \Psi^{(-)}_{ {\vec p} \, {\vec q} \, m_1 \, m_2 \, m_3 }
\mid {\hat O} \mid \Psi_{ {}^3_\Lambda H \, m } >  \mid^2
\label{e8.5}
\end{equation}
This form is convenient for the integration to determine
the total (nnp) decay rate. For the display of the angular
and energy distribution of the three nucleons, the following,
equivalent form~\cite{6} is more convenient
\[
d \, \Gamma^{\rm n+n+p} \ = \
\frac12 \, \sum_{m \, m_1 \, m_2 \, m_3} \
\mid < \Psi^{(-)}_{ {\vec p} \, {\vec q} \, m_1 \, m_2 \, m_3 }
\mid {\hat O} \mid \Psi_{ {}^3_\Lambda H \, m } >  \mid^2 \
\]
\begin{equation}
2 \pi \, d {\hat k}_1 \,  d {\hat k}_2 \,d S \,
\frac{ M_N^2 \, k_1^2 \, k_2^2 }
{\sqrt     { k_1^2 ( 2 k_2 + {\vec k}_1 \cdot {\hat k}_2 )^2  \, + \,
       k_2^2 ( 2 k_1 + {\vec k}_2 \cdot {\hat k}_1 )^2  } }
\label{e8.6}
\end{equation}
Here ${\hat k}_1$ and ${\hat k}_2$ denote the directions of two detected
nucleons. That choice of four angles relates the lab energies
$E_1$ and $E_2$ kinematically through energy and momentum conservation, 
as shown in the example
of an interparticle angle  $ \Theta_{1 2} = 180^o $
in Fig.~\ref{f1}. A point on that curve can be defined through the arclength $S$
measured from some conveniently chosen point. Our choice for $S = 0$ is shown in
Fig.~\ref{f1}. Thus instead of expressing the fivefold differential cross section
with respect to $d E_1$, for instance, we have chosen $dS$ in equation in Eq.~(\ref{e8.6}).
% This figure also illustrates how a point
%on that kinematical locus is fixed by the arclength $S$
%occuring in (\ref{e8.6}).

In (\ref{e8.6}) the necessary additional
information, whether the detected particles 1 and~2 are a proton--neutron
pair or two neutrons, has been dropped for the sake of a simple notation.
The final scattering state carries, however, additional isotopic spin
quantum numbers $\nu_1 \, \nu_2 \, \nu_3$.

Throughout the paper the normalisation of the momentum states is always like
\mbox{$ <{\vec k} \mid {\vec k}^{\ '} >
= \delta^3 ( {\vec k} - {\vec k}^{\ '} )$}.

The operator $ {\hat O} $ is of two-body character
and acts between the $\Lambda$  and a nucleon. In the hypertriton
wavefunction let us denote the  $\Lambda$ to be particle 1,
then
\begin{equation}
 {\hat O} \, = \, \sum_{i=2,3} \, {\hat O} (1,i)
\label{e18}
\end{equation}
Because of the antisymmetry of the hypertriton state and the scattering
states with respect to the two nucleons 2 and~3 the nuclear
matrix element simplifies to
\begin{equation}
< \Psi^{(-)} \mid {\hat O} \mid \Psi_{ {}^3_\Lambda H } > \, = \,
2 \, < \Psi^{(-)} \mid {\hat O} (1,2) \mid \Psi_{ {}^3_\Lambda H } >
\label{e19}
\end{equation}

The exact inclusion of the final state interactions among the 3 final nucleons
can be performed in analogy to electron scattering on ${}^3$He~\cite{9}.
We exemplify it for the nnp breakup process. For our notation in general
we refer to~\cite{10}.

The scattering state $\Psi^{(-)} \equiv  \Psi^{(-)}_{ {\vec p} \, {\vec q}}$
is Faddeev decomposed
\begin{equation}
\Psi^{(-)} \, = \, (1 + P ) \psi^{(-)}{\rm ,}
\label{e9}
\end{equation}
where $P$ is the sum of a cyclical and anticyclical permutation of 3 objects
and $ \psi^{(-)}$ is one Faddeev component.
It obeys the Faddeev equation
\begin{equation}
\psi^{(-)} \, = \, \phi^{(-)} + G_0^{(-)} t^{(-)} P  \psi^{(-)}
\label{e10}
\end{equation}
with
\begin{equation}
\phi^{(-)} \, = \, ( 1 + G_0^{(-)} t^{(-)} ) \phi^{a}_0
\label{e11}
\end{equation}
\begin{equation}
\phi^a_0 \, = \, \frac1{\sqrt{3 !}} ( 1 - P_{12} ) \mid \phi_0 > \, \equiv \,
\frac1{\sqrt{6}} ( 1 - P_{12} ) \mid {\vec p} > \mid {\vec q} >
\label{e12}
\end{equation}
Here $ G_0^{(-)}$ is the free three-nucleon  propagator,
$ t^{(-)}$ the NN (off-shell) t-matrix and $\frac1{\sqrt{6}}$
takes care of the identity of the three nucleons. Note that
$P_{12}$ acts in the two-body subsystem described by the relative momentum
$\vec p$.

Let us now insert (\ref{e9}),  (\ref{e10}) and (\ref{e11}),
into the nuclear matrix element
\[
  < \Psi^{(-)} \mid {\hat O} (1,2) \mid \Psi_{ {}^3_\Lambda H } > \, = \,
 < \psi^{(-)} \mid (1 + P) {\hat O} (1,2) \mid \Psi_{ {}^3_\Lambda H } >
\]
\[
 = \,
 < \phi^{(-)} \mid (1 + P){\hat O} (1,2) \mid \Psi_{ {}^3_\Lambda H } >  \, + \,
 < \psi^{(-)} \mid P t G_0 (1 + P){\hat O} (1,2) \mid \Psi_{ {}^3_\Lambda H } >
\]
\[
 = \, < \phi^a_0 \mid (1 + t G_0 ) (1 + P) {\hat O} (1,2)
\mid \Psi_{ {}^3_\Lambda H } >  \,
+ \, < \phi^a_0 \mid (1 + t G_0)  P t G_0 (1 + P){\hat O} (1,2)
\mid \Psi_{ {}^3_\Lambda H } >
\]
\begin{equation}
+ \, < \phi^a_0 \mid (1 + t G_0) \,  P t G_0 \,  P t G_0 (1 + P) {\hat O} (1,2)
\mid \Psi_{ {}^3_\Lambda H } > \, + \, \ldots
\label{e13}
\end{equation}
In the last equality we iterated (\ref{e10}).
It is then easily seen that this can be put into the form
\begin{equation}
<  \Psi^{(-)}_{ {\vec p} \, {\vec q}} \mid
 {\hat O} (1,2) \mid \Psi_{ {}^3_\Lambda H } > \, = \,
 < \phi^a_0 \mid (1 + P) {\hat O} (1,2) \mid \Psi_{ {}^3_\Lambda H } > \, + \,
 < \phi^a_0 \mid (1 + P) \mid U > {\rm ,}
\label{e14}
\end{equation}
where $ \mid U >$ obeys the Faddeev equation
\begin{equation}
 \mid U > \, = \, t G_0 (1 + P) {\hat O} (1,2) \mid \Psi_{ {}^3_\Lambda H } >
\, + \,  t G_0 P \mid U >
\label{e15}
\end{equation}
Note the driving term of that integral equation contains the operator
${\hat O} (1,2)$ applied to the hypertriton bound state
$ \mid \Psi_{ {}^3_\Lambda H } >$ and it also includes
rescattering terms of first order in $t$. The pure plane wave impulse
approximation is the first term on the right hand side of (\ref{e14}).

A similar reduction yields for the nd breakup process
\begin{equation}
<  \Psi^{(-)}_{ {\vec q}_0 }  \mid
 {\hat O} (1,2) \mid \Psi_{ {}^3_\Lambda H } > \, = \,
 < \phi \mid (1 + P) {\hat O} (1,2) \mid \Psi_{ {}^3_\Lambda H } > \, + \,
 < \phi \mid P \mid U > {\rm ,}
\label{e16}
\end{equation}
where now $ \mid \phi > $ contains the deuteron state $\mid \varphi_d > $:
\begin{equation}
 \mid \phi > \, = \, \mid \varphi_d > \mid {\vec q}_0 >
\label{e17}
\end{equation}
and the same state $ \mid U > $ appears.

Using the weak transition operator $ {\hat O} (1,2) $
in the context of a strictly nonrelativistic framework  requires
some approximations, which we would like to describe in the example
of pion exchange~\cite{5,7}. According to Fig.~\ref{f1.5} the
transition operator is
\begin{equation}
 {\hat O} (1,2) \, = \, i^2 \,
{\bar u} ( {{\vec k}_2}^{\ '} ) \, g_{\rm{\scriptscriptstyle NN} \pi} \,
 \gamma_5 \,
  u ( {{\vec k}_2} ) \, \frac{ F^2 ( q_\pi^2)}{q_\pi^2 - m_\pi^2 } \,
  {\bar u} ( {{\vec k}_1}^{\ '} ) \, G_F m_\pi^2 \,\,  ( A_\pi + B_\pi
\gamma_5 ) \,
  u_\Lambda ( {{\vec k}_1} )
\label{e20}
\end{equation}
Here $u$ and $ \bar u$ are the usual Dirac spinors,
$g_{\rm {\scriptscriptstyle NN} \pi}$ the strong coupling constant for the
 $NN\pi$
vertex, and $G_F m_\pi^2=2.21 \times 10^{-7}$ the weak coupling constant.
The constants $A_\pi=1.05$ and $B_\pi=-7.15$, which
determine the strength of the parity violating and parity conserving
amplitudes, respectively, have been adjusted to reproduce the decay
observables of the free $\Lambda$ particle~\cite{fldec}. We assume the
same form factor $F$ at the two vertices (the strong and the weak one).
In the nonrelativistic reduction at the weak vertex the nucleon mass,
$M_N$, and the $\Lambda$ mass, $M_\Lambda$, are replaced
by their average, $\overline M$. Then one finds
\begin{equation}
 {\hat O} (1,2) \,  \longrightarrow \, -
 G_F m_\pi^2 \,\,
 \frac {g_{\rm {\scriptscriptstyle NN} \pi}}{2 M_N} \,
\frac{ F^2 ( {\vec q}_\pi^{\ 2}) }{ {\vec q}_\pi^{\ 2} + m_\pi^2 } \,
{\vec \sigma}_2 \cdot {\vec q}_\pi \,
\left( A_\pi \, + \, \frac{B_\pi}{2 \overline{M}}
{\vec \sigma}_\Lambda \cdot {\vec q}_\pi \right)
\label{e21}
\end{equation}
with
\begin{equation}
{\vec q}_\pi \, = \,
{\vec k}_1 - {{\vec k}_1}^{\ '} \, = \, {{\vec k}_2}^{\ '} - {\vec k}_2
\label{e22}
\end{equation}

We have to use two types of Jacobi momenta, one set referring to the
hypertriton composed of $\Lambda N N $ and another set for the final state
of three nucleons. The latter ones have been already defined
in~(\ref{e4}) and will be denoted by ${\vec p}^{\ '}$ and  ${\vec q}^{\ '}$.
The ones for the hypertriton are
\[
   \vec p \, = \, \frac{ M_N {\vec k}_1 -  M_\Lambda {\vec k}_2 }
                       { M_N +  M_\Lambda }
\]
\begin{equation}
   \vec q \, = \,
\frac{ ( M_N + M_\Lambda ) {\vec k}_3 -  M_N ( {\vec k}_1 + {\vec k}_2 ) }
                       { 2 M_N +  M_\Lambda }
\label{e23}
\end{equation}
Then for total momentum zero and using the spectator condition
$ {\vec q} =  {\vec q}^{\ '} $ one has
\begin{equation}
   {\vec q}_\pi \, = \,
 {\vec p} - {\vec p}^{\ '}
\, + \,
\frac{ M_\Lambda -  M_N }{ 2 ( M_\Lambda +  M_N ) }  {\vec q}
\label{e24}
\end{equation}
As in the derivation of $ {\hat O} (1,2) $ itself we also neglect
here the difference $M_\Lambda -  M_N$
with respect to $M_\Lambda +  M_N$ and put
\begin{equation}
   {\vec q}_\pi \, \longrightarrow \,
 {\vec p} - {\vec p}^{\ '}
\label{e25}
\end{equation}
Then we get an ordinary two-body force, which does not depend
on the momentum of the third particle (which it would if the mass
difference would be included).

A final remark refers to the isospin part of the transition matrix element.
At the weak vertex the $\Lambda$ has to change into a neutron
or a proton by emission of a $\pi^0$ or $\pi^-$, respectively.
This can be formally accomplished by setting artificially the $\Lambda$
state to be
$\mid \frac12 \, -\frac12 > $
in  isospin and introducing ${\vec \tau}$
at the vertex. This is a well known trick~\cite{5,7} and is in agreement 
with the empirical $\Delta I = \frac12$ rule.
As a consequence the two-body force~(\ref{e21})
has to be multiplied by ${\vec \tau}_1 \cdot {\vec \tau}_2 $.

Now in the hypertriton the $\Lambda$-particle is treated
as a strongly interacting particle and has therefore isospin zero.
The isospin part of the hypertriton ($\Lambda$ part only) is
\begin{equation}
\mid \theta > \, = \, \mid ( \frac12 \, \frac12 ) 0 >_{23} \mid 0 0 >_1 {\rm ,}
\label{e26}
\end{equation}
where the indices denote the particles and $  ( \frac12 \, \frac12 ) 0 $
the obvious isospin coupling for the two nucleons.

Now the action of  ${\vec \tau}_1 \cdot {\vec \tau}_2 $ resulting
from the weak transition requires the $\Lambda$-particle to be treated
as $ \mid \frac12 \, -\frac12 >_1$ and consequently the isospin part
of the hypertriton has to be reinterpreted as
\[
\mid \theta > \,  \longrightarrow \mid \theta >_{\rm weak} \, \equiv \,
\mid ( \frac12 \, \frac12 ) 0 >_{23} \mid \frac12 \, -\frac12 >_1
\]
\begin{equation}
 \, = \,
\frac1{\sqrt{2}} \,
\mid \frac12 \, -\frac12 >_1 \mid \frac12 \, \frac12 >_2 \,
\mid \frac12 \, -\frac12 >_3
-\frac1{\sqrt{2}} \,
\mid \frac12 \, -\frac12 >_1 \mid \frac12 \, -\frac12 >_2 \,
\mid \frac12 \, \frac12 >_3 {\rm ,}
\label{e27}
\end{equation}
which displays the partner (nucleon~2) of the  $\Lambda$
to be a proton or a neutron, respectively.

In the nuclear matrix element one acts from the left by the isospin state
of the final three nucleon system and the  ${\vec \tau}_1 \cdot {\vec \tau}_2 $
operator and gets for the isospin part alone
\begin{equation}
{}_3 < ( t \frac12 ) T M_T \mid
{\vec \tau}_1 \cdot {\vec \tau}_2 \mid \theta >_{\rm weak} \, = \,
\delta_{M_T , -\frac12 }
\, \delta_{T , \frac12 } \, \frac{\sqrt{3}}{2} \,
\left( \sqrt{3} \delta_{t 0} + \delta_{t 1 } \right)
\label{e28}
\end{equation}
The index 3 on the bra indicates that the isospin $t$ refers to the $(1 2 )$
subsystem.

We see that only total isospin $T= \frac12$ occurs for the three final
nucleons. For isoscalar meson exchanges
${\vec \tau}_1 \cdot {\vec \tau}_2 $ is replaced by the unit
operator and the corresponding matrix element is
\begin{equation}
{}_3 < ( t \frac12 ) T M_T \mid \theta >_{\rm weak} \, = \,
\delta_{M_T , -\frac12 }
\, \delta_{T , \frac12 } \,
\left( - \delta_{t 0} + \sqrt{3} \delta_{t 1 } \right)
\label{e28.5}
\end{equation}
One can artificially separate the contributions from the proton
and neutron induced decays. This corresponds to the first and second
parts on the right hand side of (\ref{e27}), respectively.
Keeping only the one or the other both isospins $T= \frac12$
and $T=\frac32$ contribute, therefore (\ref{e28}) and (\ref{e28.5})
will be adequately modified. That separation into p--induced and n--induced
decays will be considered in section~\ref{secV}.

\section{Technicalities}
\label{secIII}

The hypertriton state contains the $ \Lambda N N $ and the $ \Sigma N N $ parts.
The $  \Lambda -  \Sigma $ conversion is crucial for the binding of the
hypertriton,
nevertheless the $ \Sigma N N $ admixture is extremely small~\cite{1}.
Thus we neglect the contribution of the  $ \Sigma$ decay and keep only the
$ \Lambda N N $ part.

In ~\cite{1} the hypertriton state has been determined in a partial wave
representation and we refer to~\cite{1} for the details of our notation.
Here we need only the form
\[
\mid \Psi_{ {}^3_\Lambda H } >  \, = \,
\sum_{\alpha} \, \int d \, p \, p^2 \,
\int d\, q \, q^2 \, \mid p q \alpha > \, \Psi_{\alpha} ( p q ) {\rm ,}
\]
where $p, q $ are the magnitudes of the Jacobi momenta (\ref{e23}) and $\alpha$
denotes the following set of discrete quantum numbers
\[
\alpha \, \equiv \,
( l s ) j \, (\lambda \frac12 ) I ( j I ) J (t \frac12 ) T
\]
Here $( l s ) j$ describe the coupling of orbital angular momentum $l$
and total spin $s$ to the total two-body angular momentum $j$ of
the $\Lambda N$ subsystem, $ (\lambda \frac12 ) I $ the corresponding
coupling of orbital and spin angular momentum of the other nucleon to its
total angular momentum $I$, $  ( j I ) J $, the resulting $jI$ coupling
to the total angular momentum $J$ and finally the isospin coupling of
$t= \frac12 $ and $\frac12$ to total isospin $T= 0$, as described above.

Also for the evaluation of the matrix elements (\ref{e14}) and~(\ref{e16})
and the solution
of the Faddeev equation~(\ref{e15}) we work in a partial wave representation,
using a complete set of basis states now for three nucleons.
They are again denoted as $  \mid p q \alpha >_N $ but
adding a subscript $N$ to indicate that the Jacobi momenta are now
from (\ref{e4}). Furthermore one has to note that this is a subset
of states antisymmetrized in the subsystem of particles 1 and~2,
thus $ ( l + s + t ) $ has to be odd.

Now projecting the Faddeev equation into the basis  $  \mid p q \alpha >_N $
and inserting appropriate decompositions of the unity one gets
\[
{}_N < p q \alpha \mid U > \, = \,
\]
\[
\sum \hspace{-0.5cm} \int \,
\sum \hspace{-0.5cm} \int  \,
{}_N < p q \alpha \mid  t G_0 (1 + P ) \mid p' q' {\alpha}' >_N \,
{}_N < p' q' {\alpha}' \mid  {\hat O} (1,2) \mid p'' q'' {\alpha}'' > \,
\Psi_{{\alpha}''} ( p'' q'') \ + \
\]
\begin{equation}
\sum \hspace{-0.5cm} \int  \,
{}_N < p q \alpha \mid  t G_0 P \mid p' q' {\alpha}' >_N \,
{}_N < p' q' {\alpha}' \mid  U >
\label{e29}
\end{equation}
This is a coupled set of integral equations, with a kernel part,
which is well known~\cite{8} from 3N scattering, and an inhomogeneous
term, whose part left of $ {\hat O} (1,2)$ is also familiar from
electron scattering~\cite{9}. What is left as a new structure
is the application of the
$ {\hat O} (1,2) $-matrix onto the wavefunction component of the
hypertriton.

Now that $ {\hat O} (1,2) $-matrix is obviously diagonal in the quantum numbers
of the spectator nucleon:
\begin{equation}
{}_N\! < p q {\alpha} \mid  {\hat O} (1,2) \mid p' q' {\alpha}' > \, = \,
\frac{ \delta ( q - q' ) }{q q' }
\, \delta_{\lambda \, {\lambda}' }
\, \delta_{I \, I' } \,
< p ( l s ) j  \mid  {\hat O} (1,2) \mid p' ( l' s' ) j >
\label{e30}
\end{equation}
and one is left with a simple  application of the two-body force onto
the hypertriton in momentum space. The right hand side of (\ref{e30})
should contain the appropriate isospin matrix element in the 3--particle
space, see (\ref{e28}) and (\ref{e28.5}), as a factor.

Once the amplitudes
$ {}_N \! < p q \alpha \mid U > $ are determined,
the matrix elements in (\ref{e14}) and (\ref{e16})
can be evaluated by quadratures in the manner described in~\cite{8}
and references therein.

\section{The transition operator}
\label{secIV}

On top of the $\pi$-induced transition potential described
in section \ref{secII} we include exchange
potentials driven by $\eta$, K, $\rho$, $\omega$ and K$^*$ mesons.
The explicit expressions for the
weak and strong Hamiltonians can be found in Ref.\cite{7}.

The resulting one boson exchange expression in a nonrelativistic
reduction for the pion is given in~(\ref{e21}) which we rewrite
here as:

\begin{equation}
V_{\pi}({\vec q}_\mu) = - G_F m_\pi^2
\frac{g}{2M_N} \left({\hat A} + \frac{\hat B}{2\overline{M}}
{\vec \sigma}_1 \cdot {\vec q}_\mu \right)
\frac{{\vec \sigma}_2 \cdot {\vec q}_\mu }{{\vec q}^{\ 2}_\mu+\mu^2}
\label{eq:pion}
\end{equation}
where $g = g_{\rm {\scriptscriptstyle NN} \pi}$
is the strong coupling constant for the NN$\pi$ vertex, $\mu$
is the pion mass, ${\vec q}_\mu$ stands now for the momentum carried
by the exchanged meson and the operators
${\hat A}$ and ${\hat B}$ contain the isospin
dependence of the potential

\begin{eqnarray}
{\hat A} &=& A_\pi \,\, {\vec \tau}_1 \cdot {\vec \tau}_2 \nonumber \\
{\hat B} &=& B_\pi \,\, {\vec \tau}_1 \cdot {\vec \tau}_2 \ .
\end{eqnarray}

For pseudoscalar mesons different from the pion we have an expression
analogous to (\ref{eq:pion})
but making the following replacements:
\begin{eqnarray}
g &\to& \,g_{\rm {\scriptscriptstyle NN} \eta}, \nonumber \\
\mu &\to& m_\eta, \nonumber \\
{\hat A} &\to& A_\eta, \nonumber \\
{\hat B} &\to& B_\eta .
\label{eq:eta}
\end{eqnarray}
when considering the exchange of the isoscalar $\eta$-meson, and
\begin{eqnarray}
g &\to& g_{\rm {\scriptscriptstyle \Lambda NK}}, \nonumber \\
\mu &\to& m_{\rm {\scriptscriptstyle K}}, \nonumber \\
{\hat A} &\to& \left( \frac{
C^{\rm \scriptscriptstyle PV}_{\rm \scriptscriptstyle K}}{2} +
D^{\rm \scriptscriptstyle PV}_{\rm \scriptscriptstyle K} + \frac{
C^{\rm \scriptscriptstyle PV}_{\rm \scriptscriptstyle K}} {2}
{\vec \tau}_1 \cdot
{\vec \tau}_2 \right) \frac{M_N}{\overline M} \nonumber \\
{\hat B} &\to& \left( \frac{
C^{\rm \scriptscriptstyle PC}_{\rm \scriptscriptstyle K}} {2} +
D^{\rm \scriptscriptstyle PC}_{\rm \scriptscriptstyle K} + \frac{
C^{\rm \scriptscriptstyle PC}_{\rm \scriptscriptstyle K}} {2}
{\vec \tau}_1 \cdot
{\vec \tau}_2 \right)
\label{eq:kaon}
\end{eqnarray}
for the isodoublet kaon.
%, where the explicit values of the strong
%and weak coupling constants appearing in~(\ref{eq:eta})
%and (\ref{eq:kaon}) are given in Ref.~\cite{parre}.

The factor $M_N / \overline M$
corrects for the fact that the nonrelativistic reduction of the
strong $\Lambda NK$ vertex is now proportional to
$\frac{{\vec \sigma}_2 \cdot {\vec q}_\mu}{ 2 \overline M} $,
giving a factor $1 / \overline M$ instead of
$1 / M_N$.

In the case of vector mesons as the $\rho$, one
obtains\cite{7}:

\begin{eqnarray}
{V_{\rho}}({\vec q}_\mu)  &=&
G_F m_\pi^2
 \left( F_1 {\hat \alpha} - \frac{({\hat \alpha} + {\hat \beta} )
 ( F_1 + F_2 )} {4M_N \overline{M}}
({\vec \sigma}_1 \times {\vec q}_\mu)
({\vec \sigma}_2 \times {\vec q}_\mu) \right. \nonumber \\
& & \phantom { G_F m_\pi^2 A }
\left. +i \frac{{\hat \varepsilon} ( F_1 + F_2 )} {2M_N}
({\vec \sigma}_1 \times
{\vec \sigma}_2 ) \cdot {\vec q}_\mu \right)
\frac{1}{{\vec q}^{\ 2}_\mu + \mu^2} \
\label{eq:rhopot}
\end{eqnarray}
with $\mu = m_\rho$, $F_1 = g^{\rm \scriptscriptstyle V}_{\rm {\scriptscriptstyle NN} \rho}$,
$F_2 = g^{\rm \scriptscriptstyle T}_{\rm {\scriptscriptstyle NN} \rho}$
and where the operators ${\hat \alpha}$, ${\hat
\beta}$ and ${\hat \varepsilon}$ have the following structure:

\begin{eqnarray}
{\hat \alpha} &=& \alpha_\rho \,\, {\vec \tau}_1 \cdot
{\vec \tau}_2  \nonumber \\
{\hat \beta} &=& \beta_\rho \,\,
{\vec \tau}_1 \cdot {\vec \tau}_2
\nonumber \\
{\hat \varepsilon} &=& \varepsilon_\rho \,\,
{\vec \tau}_1 \cdot {\vec \tau}_2
\end{eqnarray}

We can get the nonrelativistic potential corresponding to the
exchange of the rest of vector mesons by making the following
replacements in Eq. (\ref{eq:rhopot}):

\begin{eqnarray}
\mu &\to& m_\omega, \nonumber \\
F_1 &\to& g^{\rm \scriptscriptstyle V}_{\rm {\scriptscriptstyle NN} \omega}, \nonumber \\
F_2 &\to& g^{\rm \scriptscriptstyle T}_{\rm {\scriptscriptstyle NN} \omega}, \nonumber \\
{\hat \alpha} &\to& \alpha_\omega, \nonumber \\
{\hat \beta} &\to& \beta_\omega, \nonumber \\
{\hat \varepsilon} &\to& \varepsilon_\omega.
\end{eqnarray}
for the exchange of the isoscalar $\omega$, and
\begin{eqnarray}
\mu &\to& m_{\rm {\scriptscriptstyle K^*}}, \nonumber \\
F_1 &\to& g^{\rm \scriptscriptstyle V}_{\rm {\scriptscriptstyle \Lambda N K^*}}, \nonumber \\
F_2 &\to& g^{\rm \scriptscriptstyle T}_{\rm {\scriptscriptstyle \Lambda N K^*}}, \nonumber \\
{\hat \alpha} &\to& \frac{
C^{\rm \scriptscriptstyle PC, V}_{\rm \scriptscriptstyle K^*}}{2} +
D^{\rm \scriptscriptstyle PC, V}_{\rm \scriptscriptstyle K^*} + \frac{
C^{\rm \scriptscriptstyle PC, V}_{\rm \scriptscriptstyle K^*}}{2}
{\vec \tau}_1 \cdot {\vec \tau}_2,
\nonumber \\
{\hat \beta} &\to& \frac{
C^{\rm \scriptscriptstyle PC, T}_{\rm \scriptscriptstyle K^*}}{2} +
D^{\rm \scriptscriptstyle PC, T}_{\rm \scriptscriptstyle K^*} +  \frac{
C^{\rm \scriptscriptstyle PC, T}_{\rm \scriptscriptstyle K^*}}{2}
{\vec \tau}_1 \cdot {\vec \tau}_2,
\nonumber \\
{\hat \varepsilon} &\to& \left( \frac {
C^{\rm \scriptscriptstyle PV}_{\rm \scriptscriptstyle K^*}}{2} +
D^{\rm \scriptscriptstyle PV}_{\rm \scriptscriptstyle K^*} + \frac{
C^{\rm \scriptscriptstyle PV}_{\rm \scriptscriptstyle K^*}}{2}
{\vec \tau}_1 \cdot {\vec \tau}_2
\right) \frac{M_N}{\overline M}
\label{eq:kstar}
\end{eqnarray}
for the isodoublet $K^*$-meson.

In configuration space the potential including the exchange of all
the mesons can be cast into the form:

\begin{eqnarray}
V({\vec r}) &=& \sum_{i} \sum_\alpha V_\alpha^{(i)}
({\vec r}) = \sum_i \sum_{\alpha}
V_\alpha^{(i)} (r) \hat{O}_\alpha \hat{I}_\alpha^{(i)} \nonumber
\\
&=& \sum_{i} \left[ V_C^{(i)}(r) \hat{I}^{(i)}_C + V_{SS}^{(i)}(r)
{\vec \sigma}_1 \cdot {\vec \sigma}_2
%\mbox{\boldmath $\sigma$}_1 \mbox{\boldmath $\sigma$}_2
\hat{I}^{(i)}_{SS}
+ V_T^{(i)}(r)
S_{12}(\hat{r}) \hat{I}^{(i)}_T + \right. \nonumber \\
& & \left. + \left( n^i {\vec \sigma}_2 \cdot {\hat r}
%\mbox{\boldmath $\sigma$}_2
%\cdot \hat{\bf r}
+ (1-n^i) \left[ {\vec \sigma}_1 \times {\vec \sigma}_2
%\mbox{\boldmath $\sigma$}_1 \times
%\mbox{\boldmath $\sigma$}_2
\right] \cdot \hat{r} \right)
V_{PV}^{(i)}(r) \hat{I}^{(i)}_{PV} \right] \ ,
\label{eq:genpot}
\end{eqnarray}
where the index $i$ runs over the different mesons exchanged ($i=1,\dots,
6$ meaning $\pi,\rho$,K,K$^*$,$\eta,\omega$)
and $\alpha$ over the different
spin operators denoted by $C$
(central spin independent), $SS$ (central spin dependent), $T$
(tensor) and $PV$ (parity violating). In the above expression,
particle~1 refers to the
$\Lambda$ and
$n^i = 1 (0)$ for pseudoscalar (vector) mesons.
For isovector mesons ($\pi$, $\rho$) the isospin factor $\hat{I}^{(i)}_\alpha$
is ${\vec \tau}_1 \cdot {\vec \tau}_2$ for all values of $\alpha$,
%$\mbox{\boldmath $\tau$}_1
%\mbox{\boldmath $\tau$}_2$,
for isoscalar mesons ($\eta$,$\omega$)
this factor is just $\hat{1}$,
and for isodoublet mesons (K,K$^*$) there
are contributions proportional to $\hat{1}$ and to
${\vec \tau}_1 \cdot {\vec \tau}_2$
%$\mbox{\boldmath $\tau$}_1 \mbox{\boldmath $\tau$}_2$
with coefficients that depend on the
coupling constants and, therefore, on the
spin structure piece of the potential denoted by $\alpha$.

For K-exchange the isospin factors in
(\ref{eq:genpot}) are

\begin{eqnarray}
{\hat I}_{C}^{(3)}&=& 0 \nonumber \\
{\hat I}_{SS}^{(3)}&=& {\hat I}_{T}^{(3)} =
\frac{
C^{\rm \scriptscriptstyle{P C}}_{\rm\scriptscriptstyle{K}}} {2} +
D^{\rm \scriptscriptstyle{P C}}_{\rm\scriptscriptstyle{K}} + \frac{
C^{\rm \scriptscriptstyle{P C}}_{\rm\scriptscriptstyle{K}}} {2}
%\mbox{\boldmath $\tau$}_1 \mbox{\boldmath $\tau$}_2
{\vec \tau}_1 \cdot {\vec \tau}_2 \nonumber \\
{\hat I}_{PV}^{(3)} &=&
\frac{
C^{\rm \scriptscriptstyle{P V}}_{\rm\scriptscriptstyle{K}}} {2} +
D^{\rm \scriptscriptstyle{P V}}_{\rm\scriptscriptstyle{K}} + \frac{
C^{\rm \scriptscriptstyle{P V}}_{\rm\scriptscriptstyle{K}}} {2}
%\mbox{\boldmath $\tau$}_1 \mbox{\boldmath $\tau$}_2
{\vec \tau}_1 \cdot {\vec \tau}_2
\end{eqnarray}
and for K$^*$ exchange they are 
\begin{eqnarray}
{\hat I}_{C}^{(6)} &=&  \frac{
C^{\rm \scriptscriptstyle{PC,V}}_{\rm\scriptscriptstyle{K^*}}} {2} +
D^{\rm \scriptscriptstyle{PC,V}}_{\rm\scriptscriptstyle{K^*}} + \frac{
C^{\rm \scriptscriptstyle{PC,V}}_{\rm\scriptscriptstyle{K^*}}} {2}
%\mbox{\boldmath $\tau$}_1 \mbox{\boldmath $\tau$}_2
{\vec \tau}_1 \cdot {\vec \tau}_2 \nonumber \\
{\hat I}_{SS}^{(6)} &=& {\hat I}_{T}^{(6)} =
\frac { \left(
C^{\rm \scriptscriptstyle{PC,V}}_{\rm\scriptscriptstyle{K^*}} +
C^{\rm \scriptscriptstyle{PC,T}}_{\rm\scriptscriptstyle{K^*}} \right)} {2}
+ \left( D^{\rm \scriptscriptstyle{PC,V}}_{\rm\scriptscriptstyle{K^*}} +
D^{\rm \scriptscriptstyle{PC,T}}_{\rm\scriptscriptstyle{K^*}} \right) +
\frac{ \left( C^{\rm \scriptscriptstyle{PC,V}}_{\rm\scriptscriptstyle{K^*}} +
C^{\rm \scriptscriptstyle{PC,T}}_{\rm\scriptscriptstyle{K^*}} \right) }{2}
%\mbox{\boldmath $\tau$}_1 \mbox{\boldmath $\tau$}_2
{\vec \tau}_1 \cdot {\vec \tau}_2 \nonumber \\
{\hat I}_{PV}^{(6)} &=&
\frac{
C^{\rm \scriptscriptstyle{PV}}_{\rm\scriptscriptstyle{K^*}}} {2} +
D^{\rm \scriptscriptstyle{PV}}_{\rm\scriptscriptstyle{K^*}} + \frac {
C^{\rm \scriptscriptstyle{PV}}_{\rm\scriptscriptstyle{K^*}}} {2}
%\mbox{\boldmath $\tau$}_1  \mbox{\boldmath $\tau$}_2
{\vec \tau}_1 \cdot {\vec \tau}_2 \ .
\end{eqnarray}

The different pieces $V_{\alpha}^{(i)}$, with $\alpha=C,SS,T,PV$,
given in Ref.\cite{7}, are
reproduced here for completeness
\begin{eqnarray}
V_{C}^{(i)} (r) &=&  K^{(i)}_{C}
\frac {{\rm e}^{- \mu_i r}} {4 \pi r} \equiv K^{(i)}_{C} \:
V_{C} (r,\mu_i)
\label{eq:cpot}
 \\
V_{SS}^{(i)}(r) &=& K^{(i)}_{SS}  \frac{1}{3}
\: \left[ {\mu_i}^2  \: \frac {{\rm e}^{- \mu_i r}} {4 \pi r}
- \delta (r) \right] \equiv K^{(i)}_{SS} V_{SS} (r,\mu_i)
\label{eq:sspot} \\
V^{(i)}_{T} (r) &=& K^{(i)}_{T} \: \frac {1}{3}
\:  \mu_i^2  \: \frac {{\rm e}^{- \mu_i r}} {4 \pi r} \:
\left( 1 + \frac{3} {{\mu_i} r} + \frac{3} {({\mu_i
r})^2} \right)
 \equiv K^{(i)}_{T} V_{T} (r,\mu_i)
\label{eq:tpot}
 \\
V_{PV}^{(i)}(r) &=& K^{(i)}_{PV} \: \mu_i \:
\frac {{\rm e}^{- {\mu_i} r}} {4 \pi r}
\left( 1 + \frac{1} {\mu_i r} \right)
 \equiv  K^{(i)}_{PV} V_{PV}(r,\mu_i)  \ .
\label{eq:pvpot}
\end{eqnarray}
where $\mu_i$ denotes the mass of the
different mesons.
The expressions for $K^{(i)}_\alpha$, which contain factors and coupling
constants, are given in Table~\ref{tab:k}. The explicit values of the
strong and weak coupling constants, taken from Ref. \cite{7}, are
shown in Table~\ref{tab:const}.

Including monopole form factors
$F_{i}({\vec q}^{\ 2})=
(\Lambda_i^2-\mu_i^2)/(\Lambda_i^2+{\vec q}^{\ 2})$
at both vertices, where the value of the cut-off,
$\Lambda_i$, depends on the meson (see Table \ref{tab:const}),
leads to the following regularization of the potential:

\begin{eqnarray}
V_{C} (r; \mu_i) &\to& V_{C} (r; \mu_i) - V_{C}
(r;\Lambda_i) -
\Lambda_i \frac{ {\Lambda_i}^2 - {\mu_i}^2}{2} \frac{{\rm e}^{-
\Lambda_i r}}{4 \pi}
\left( 1 - \frac{2}{\Lambda_i r} \right)  \\
V_{SS} (r; \mu_i) &\to& V_{SS} (r; \mu_i) - V_{SS}
(r;\Lambda_i) -
\Lambda_i \frac{ {\Lambda_i}^2 - {\mu_i}^2}{2} \frac{{\rm e}^{-
\Lambda_i r}}{4 \pi}
\left( 1 - \frac{2}{\Lambda_i r} \right)  \\
V_{ T} (r; \mu_i) &\to& V_{ T} (r; \mu_i) - V_{T}
(r; \Lambda_i) -
\Lambda_i \frac{ {\Lambda_i}^2 - {\mu_i}^2}{2} \frac{{\rm e}^{-
\Lambda_i r}}{4 \pi}
\left( 1 + \frac{1}{\Lambda_i r} \right)  \\
V_{PV} (r; \mu_i) &\to& V_{PV} (r; \mu_i) - V_{PV}
(r; \Lambda_i) -
\frac{ {\Lambda_i}^2 - {\mu_i}^2}{2} \frac{{\rm e}^{- \Lambda_i r}}
{4\pi}
\end{eqnarray}
where $V_{\alpha} (r; \Lambda_i)$ has the same structure as
$V_{\alpha}(r;\mu_i)$, defined in Eqs.
(\ref{eq:cpot})--(\ref{eq:pvpot}), but replacing the meson mass
$\mu_i$ by the corresponding cutoff mass $\Lambda_i$.

The last step is the transition into the momentum space partial wave
representation (\ref{e30}). Of course we could have derived that directly
from (\ref{eq:pion}--\ref{eq:rhopot}) using the standard helicity
formalism~\cite{11}.

This leads to
\[
< p ( l s ) j \mid V^{(i)} \mid p' ( l' s' ) j > \ = \
\frac2{\pi} \, i^{(l'-l)} \,
\int_0^\infty \, d\, r \, r^2 j_l ( p r )\,  V_C^{(i)} ( r )  \,
j_{l'} ( p' r ) \, \delta_{l \, l'} \, \delta_{s \, s'} \ + \
\]
\[
\frac2{\pi} \, i^{(l'-l)} \,
\int_0^\infty \, d\, r \, r^2 j_l ( p r ) \, V_{\rm SS}^{(i)} ( r ) \,
j_{l'} ( p' r ) \,
< ( l s ) j \mid {\vec \sigma}_1 \cdot {\vec \sigma}_2 \mid ( l' s' ) j > \ + \
\]
\[
\frac2{\pi} \, i^{(l'-l)} \,
\int_0^\infty \, d\, r \, r^2 j_l ( p r ) V_T^{(i)} ( r ) \, j_{l'}
( p'r )
\,
< ( l s ) j \mid S_{12} ( \hat r ) \mid ( l' s' ) j > \ + \
\]
\[
\frac2{\pi} \, i^{(l'-l)} \, n^i \,
\int_0^\infty \, d\, r \, r^2 j_l ( p r ) V_{\rm PV}^{(i)} ( r ) \,
j_{l'}
(p' r ) \,
< ( l s ) j \mid  {\vec \sigma}_2 \cdot {\hat r}  \mid ( l' s' ) j >
\ + \
\]
\begin{equation}
\frac2{\pi} \, i^{(l'-l)} \, (1 - n^i) \,
\int_0^\infty \, d\, r \, r^2 j_l ( p r ) V_{\rm PV}^{(i)} ( r ) \,
j_{l'}
(p' r ) \,
< ( l s ) j \mid [{\vec \sigma}_1 \times {\vec \sigma}_2] \cdot {\hat r}
\mid ( l' s') j >
\label{e53}
\end{equation}
The radial integrations were carried out numerically. The angular momentum
parts are standard and are given for instance in~\cite{7}.

Since our results show a strong dependence on the different meson
contributions with varying signs, we would like to display the radial
shapes of the four types of potentials $(C,\ SS,\ T,\ PV)$
split into the different
meson contributions.
This is shown in Figs. \ref{fig:c}a and \ref{fig:c}b for the central
spin-independent, in Figs. \ref{fig:ss}a and \ref{fig:ss}b for
the central spin-dependent, in Figs. \ref{fig:t}a and \ref{fig:t}b
for the tensor and in Figs. \ref{fig:pv}a and \ref{fig:pv}b for the
parity violating channels. Note that we have represented
$r^2 V^{(i)}_\alpha$, where $V^{(i)}_\alpha$ is the potential
regularized
by the effect of form factors, and that the expectation value of the
isospin factor for each meson and channel has also been included.
As expected, we observe that the $\pi$-meson contribution is by
far the one of longest range. More interesting is to note that,
compared to the pion, all the other mesons play a relevant role
in a wide range which extends up to about $1.5$ fm.
On the right hand side of the figures we
have plotted the full potential obtained when all meson contributions
are added. We observe that the full potential is clearly different from
the $\pi$-only one, shown on the left hand side of the figures. We
also see that in the spin independent central channel (Figs.~\ref{fig:c}a,b)
the contribution of the vector mesons compensate each other giving
rise to a practically negligible transition potential for both
isospin channels.

In Figs. \ref{fig:c}a,b only the vector mesons appear since they are
the ones that contribute to the spin-independent channel.
These figures show that the three potentials have about the same range
and their contribution is similarly relevant.

As seen in Figs. \ref{fig:ss}a,b, the K$^*$ meson gives
a very important contribution to the
central spin-dependent channel.
We also observe that, except in the intermediate range where
the potentials change sign, there is a constructive
interference between the pseudoscalar and vector components
of each isospin-like pair
[ ($\pi,\rho$) , (K,K$^*$) , ($\eta,\omega$) ].
Note that, in the $T=1$ channel, the $\omega$-meson lies very close to the
$\rho$-meson potential. This is due to the similar value of the
$\rho$ and $\omega$ masses and to the fact that the combination
of strong and weak coupling constants building up 
$K^{(i)}_{\rm {\scriptscriptstyle
SS}}$ (see Tables \ref{tab:k} and \ref{tab:const} ) gives, 
by chance, a very similar value. This behavior
is not observed for the $T=0$ channel because, due to its isovector
character, the $\rho$-meson contains an additional factor of $-3$
compared to the $\omega$-meson, as can be clearly seen in
Fig.\ref{fig:ss}a.

The tensor transition potential is shown in Figs \ref{fig:t}a,b.
In this case, we observe a destructive
interference pattern for each pair of isospin-like mesons.
In the $T=1$ channel the K$^*$-meson clearly stands out with respect to
the other mesons and, for the same reasons explained above, the
$\rho$ and $\omega$ contributions are again very similar.

Figs. \ref{fig:pv}a,b show the parity violating contributions.
Here, the interference is constructive for the ($\eta,\omega$) pair
and destructive for the ($\pi,\rho$) pair in both isospin channels. The 
(K,K$^*$) pair shows a destructive interference in the $T=0$ channel
and a constructive one in the $T=1$ channel. In these
figures the longest range of the pion contribution stands out
quite clearly over the other mesons, especially in the $T=0$
channel.

\section{Results}
\label{secV}

We used a hypertriton wave function based on the Nijmegen~93 NN
potential~\cite{12}
and the Nijmegen YN interaction~\cite{2}, which include
the $\Lambda - \Sigma$
transitions. The number of channels (see section~\ref{secIII}) used in the
solution of the corresponding Faddeev equation is 102. This leads to a fully
converged state, which has the proper antisymmetrisation among the
two nucleons built in. Also the NN and YN correlations are
exactly included  as generated by the various baryon-baryon forces
(see~\cite{1}). The $\Sigma NN$ part of the state has a probability
of 0.5 \% and will be neglected.

The deuteron and the final state interaction among the three nucleons
is generated using the Nijmegen~93 NN force, including
the NN force components up to total two-body angular
momentum $j = 2$. This is sufficient to get a converged result for the
nuclear matrix element.

Since the total three-body angular momentum is conserved, the Faddeev
equation (\ref{e15}) for the final state interaction (FSI)
has to be solved only for total
three-body angular momentum $ J= \frac12$,
but for both parities due to the parity violating transition potential.

The total nonmesonic
decay rate turns out to be $\Gamma = 0.21 \times 10^8  {1 \over s} $,
which is 0.55~\% of the free $\Lambda$ decay rate,
$\Gamma_{\Lambda} = 3.8 \times 10^9  {1 \over s} $.
This is about one order of magnitude smaller
than what has
been found in the rough  estimate~\cite{4}, which was based only
on $\pi$-exchange, a simplified hypertriton wave function and the
absence of FSI.
In the following, we show that the final value for the total decay rate
comes from many
dynamical ingredients, which all contribute significantly. Therefore,
that quantity will be an important test for our understanding of that system
and should be measured.

Table~\ref{t1} shows the individual contributions of the six mesons
to $\Gamma_{nm}$ and the way each meson contributes to the final
result.
We see that the pion by itself provides the largest contribution, followed
by K, K$^*$ and $\omega$. Adding the meson contributions one by one can yield
a strongly varying sequence as seen in Table~\ref{t1} choosing a special
but arbitrary order.
In view of Figs.~\ref{fig:c}--\ref{fig:pv} this is hardly surprising.
The final result, however, is such that one ends up close to the value with
pion exchange only.

The total decay rate is the sum of the partial rates for the nd and nnp
decays. Our result for the pion only are shown in Table~\ref{t2it3}.
The nnp contribution is clearly dominant. We also show the plane wave impulse 
approximation (symmetrized) (PWIAS) results. They are defined by evaluating 
the nuclear matrix elements in Eqs.~(\ref{e14}) and (\ref{e16}) keeping only
the first terms, respectively.
The comparison of PWIAS to the full result (keeping both terms in Eqs.~(\ref{e14}) 
and (\ref{e16})) underlines the importance of the final 
state interaction, which reduces the rate.
Finally the parity conserving and parity violating contributions are listed
and it is seen that they are comparable to each other, though with a slight
dominance of the parity conserving part.

The corresponding numbers including all mesons are also displayed in Table~\ref{t2it3}.
Again the final state interaction is very important and reduces the PWIAS
results by about a factor of 2. Now for all mesons included the parity
conserving part is clearly dominant.

There is often a separation of p-- and n--induced decay in the
literature~\cite{5,7}. They act clearly coherently and strictly spoken
cannot be separated experimentally.
Theoretically, however,
we can choose in the intermediate state
in front of $\mid  \Psi_{ {}^3_\Lambda H } >$ in Eqs.~(\ref{e14}),
(\ref{e15}) and (\ref{e16}) a situation that
the $\Lambda$-particle chooses either a proton or a neutron
as its partner for meson exchanges.
This amounts to keeping only the first or second term on the right hand
side of Eq.~(\ref{e27}), respectively.
As already mentioned above this requires to keep also $T=3/2 $
states in the final state.

Our results for the pion only and for all 6 mesons
are displayed in Table~\ref{t4it5}.
For the nd breakup clearly the separate n-- and p--induced decay rates do not
add up to the total nd--decay rate, which tells that there is interference.
On the other hand for the nnp breakup the separate contributions
from the n-- and p-induced decays add up to the total nnp--decay rate.
However, this does not imply automatically that they can be separated
experimentally. We come back to that interesting issue below.
We see that for nd and nnp decay the p--induced decay is stronger.

Focusing on the PWIAS for the nnp breakup of Eq.~(\ref{e14}) one has
three contributions
\[
 < \phi^a_0 \mid (1 + P) {\hat O} (1,2) \mid \Psi_{ {}^3_\Lambda H } > \, = \,
\frac1{\sqrt{6}} \,
{}_{12}^{\ a} \! < {\vec p} m_1 m_2 \, \nu_1 \nu_2
\mid {}_3 \! < {\vec q} m_3 \nu_3
\mid {\hat O} (1,2)  \mid \Psi_{ {}^3_\Lambda H } > \,
\]
\[
 + \, \frac1{\sqrt{6}} \,
{}_{23}^{\ a} \! < {\vec p} m_1 m_2 \, \nu_1 \nu_2
\mid {}_1 \! < {\vec q} m_3 \nu_3
\mid {\hat O} (1,2)  \mid \Psi_{ {}^3_\Lambda H } >
\]
\begin{equation}
\ \ \, + \,
\frac1{\sqrt{6}} \,
{}_{31}^{\ a} \! < {\vec p} m_1 m_2 \, \nu_1 \nu_2
\mid {}_2 \! < {\vec q} m_3 \nu_3
\mid {\hat O} (1,2)  \mid \Psi_{ {}^3_\Lambda H } >
\label{e54}
\end{equation}
In the above equation
 we applied the $P$ operator to the left. As always, the subscripts
refer to the particles in states with momenta
$\vec p$, $\vec q$, spin magnetic quantum numbers $m$ and neutron or proton
labels $\nu$. Using Eq.~(\ref{e27}) the isospin matrix elements can
simply be calculated with the result that in the first matrix element on the
right hand side of Eq.~(\ref{e54}) nucleons $1$ and $2$ are two neutrons
for n--induced decay and a neutron--proton pair for the p--induced decay.
Also that first matrix element peaks at $ \vec q = 0 $, which means
that nucleons $1$ and $2$ share the total energy and emerge back to back.
Under this kinematical condition the other two matrix elements are
strongly suppressed, as is manifest if one expresses the momenta
occurring in these two matrix elements in terms of $\vec p$ and $\vec q = 0$
of the first matrix element.
If we denote the $\vec p$ from the first matrix element as ${\vec p}_{12}$,
then it turns out that
$\vec p = - \frac12 {\vec p}_{12} (  - \frac12 {\vec p}_{12}) $ and
$\vec q = {\vec p}_{12} (  -{\vec p}_{12}) $ in the second (third)
matrix element, respectively, and for such a $\vec q$-value
$ \mid \Psi_{ {}^3_\Lambda H } > $ is suppressed.

The other two matrix elements also peak if particles 1 or~2 emerge with zero
momenta. Therefore, we have to expect 3 peaks. Let us now take a closer
look at the quantity $ d \Gamma^{n + n +p} / dp d \theta$, defined
in Eq.~\ref{e8.5}. That quantity,
suitably restricted to certain or all meson exchanges,
summed over all $p$ and $\theta$ values
provided the various values of Tables~\ref{t1}-\ref{t4it5}.
For the choice of Jacobi
momenta~(\ref{e4}) the 3 peaks are located as sketched in Fig.~\ref{f2}.
Energy and momentum conservation requires that 
$ p^2 + \frac34 q^2 \equiv p_{\rm max}^2$, where $p_{\rm max}$
is the maximal $p$-value.
For the available energy $p_{\rm max} \approx 2 \, {\rm fm}^{-1}$.
As an example we regard ${\vec k}_2 = 0 $. Then
$ {\vec p} = \frac12 {\vec k}_1 $ and 
$ {\vec q} = -{\vec k}_1 $.
It follows that $ k_1 = p_{\rm max}$ and consequently 
$p = \frac12 p_{\rm max}$.
For ${\vec k}_3 = 0$ the momenta ${\vec k}_1$ and ${\vec k}_2$ have to be
back to back and there will be no $\theta$ dependence;
for ${\vec k}_2 = 0$ the momenta ${\vec k}_3$ and ${\vec k}_1$ are
opposite to each other, therefore $\theta$ = $\pi$ and finally for
${\vec k}_1 = 0$ the momenta ${\vec k}_3 = -{\vec k}_2$ and
$\theta = 0$.
For PWIAS the result is displayed in Fig.~\ref{f4} for
the exchange of all mesons. The corresponding results for $\pi$ exchange only 
is qualitatively similar but larger by about 50 \%.
As expected we see the 3 peaks at the proper locations.
The variation with $\theta $ for the maximal $p$-value is due to
the factor $\sin \theta$ in the expression (\ref{e8.5}).
Note that the $\sin \theta$-dependence also removes the highest peak values for
 all
3 peaks.

For the choice of nucleons 1 and 3 being neutrons and 2 a proton we have
thus to expect that for a neutron induced decay there should be only one peak
at the position $ p= \frac12 p_{\rm max}$ and $\theta = \pi$, which is indeed
the case as shown
in Fig.~\ref{f5}. Note that in evaluating the nuclear matrix element 
of Eq.~\ref{e8.5} we fixed the isospin magnetic quantum numbers 
$\nu_1$, $\nu_2$, $\nu_3$ to be $-\frac12$, $\frac12$, $-\frac12 $.
This refers to all figures \ref{f4}--\ref{f10}. 
For p-induced decay we expect two peaks
corresponding to either ${\vec k}_1 = 0$ or ${\vec k}_2 = 0$.
And this is what comes out and what is shown in Fig.~\ref{f6}.
Regarding Figs.~\ref{f5}--\ref{f6} we see that the areas populated
by n--  and p--induced decays appear to be well separated in phase space
and seem to add up
essentially incoherently to the full result. A closer inspection, however,
will be carried through below, which leads to a different result.

Now let us turn on the final state interaction. For 3 nucleons
interacting among each other one knows from Nd breakup reactions~\cite{8},
that cross sections are strongly enhanced if two nucleons emerge with
equal momenta. This is due to the strong interaction in the ${}^1 S_0$
state, where the NN t-matrix has a pole close to zero
energy (virtual state). These enhancements will be called final state
interaction peaks (FSIP) in the following.
Fig.~\ref{f2} shows the positions, where this happens in the
$p - \theta $ plane. For the case $p = 0$ clearly
no $\theta$-dependence is present.
For all meson exchanges
Fig.~\ref{f8}
shows $d \Gamma^{\rm n+n+p} / dp d \theta$ 
including the full final state interaction.
We see again
the 3 peaks already known from the PWIAS result, but
with reduced heights according to the already known reduction of the rate due
to the final state interaction. In addition there are two more little peaks
caused by the final state interaction for two pairs of nucleons,
where the nucleons forming a pair have equal momenta, respectively.
The final state interaction peak for the third pair is suppressed
by the kinematical factor $p^2$ in~(\ref{e8.5}).
The p-- and n--induced pictures (Figs.~\ref{f9}--\ref{f10})
keeping the full final state interaction
again look qualitatively similar to the ones evaluated in the PWIAS
approximation, only the final state interaction peaks are added.

It is interesting to see despite the fact that FSI decreases
the $d \Gamma^{\rm n+n+p} / dp d \theta$ values significantly, which means
a strong rescattering among the three nucleons, that there is only one peak
for the n--induced decay and the rescattering does not populate the other
two peak areas. Corresponding is also true for the p--induced decay.
Again the events for n-- and p--induced decays seem to add up incoherently in
the quantity $d \Gamma^{\rm n+n+p} / dp d \theta$.

Let us now discuss the form (\ref{e8.6}) of the differential
decay rate expressed in individual momenta in the total
momentum zero frame of the decaying hypertriton. After
averaging over the initial state polarization and summing
over the spin magnetic quantum numbers of the final 3 nucleons
the decay rate can depend only on the angle between the
two nucleon detectors, $ \Theta_{1 2} $. We now show two sets of figures,
Figs.~\ref{f11}--\ref{f12},
for various $\Theta_{1 2}$'s $ 0^o \le \Theta_{1 2} \le 180^o $
and for pion exchange only. In Fig.~\ref{f11} we compare PWIAS to the
full calculation including final state interaction. Thereby the
two detected nucleons can be either a proton neutron pair or two
neutrons. For $ \Theta_{12} =180^o $ we see a strong enhancement
for $S \approx 110 $ MeV.  (This corresponds to a location around 
the middle of the locus in Fig.~\ref{f1}.)
In PWIAS this is caused by the fact that there
the ${}^3_\Lambda$H wavefunction enters at $\vec q = \vec k_3 = 0$.
Like in Nd scattering
we shall from now on call such a configuration, where one final
nucleon has zero momentum, a quasi free scattering (QFS) configuration.
The final state interaction reduces that enhancement, but
it is still
pronounced. In addition we see two FSIP's
in the full calculation. (They have to be absent of course in PWIAS).
For $\Theta_{12}=160^o$ that enhancement is reduced and two peaks emerge
at the beginning  and the end of the S-curve. Since there either $ E_1 $
or $ E_2$ are small, we have again configurations, which are close to
QFS conditions, now for the nucleon pairs 2,3 and 1,3, respectively.
This explains the additional enhancements.
Now at $\Theta_{12} =120^o $ the enhancement
in the middle of the S-curve has disappeared.
That point on the S-curve corresponds exactly
to the so called space-star configuration in a $ N+d \rightarrow N+N+N $
process. All three nucleons receive the same energy and emerge completely
symmetrically under $120^o$ pairwise angles. This is far away from QFS
 conditions
and no enhancement is expected. Of course, at the beginning and end
of the S-curve the peaks correspond again to conditions close to QFS.
The situation remains similar at $\Theta_{12}=90^o $ and $ 60^o $. Finally
a new structure appears at $\Theta_{12} = 20^o$  and above all at
 $\Theta_{12}=0^o$
in the middle of the S-curve. This is a FSIP,
which is fully developed for $\Theta_{12}=0^o$.
 In principle it could be used to extract
information on the np and nn scattering lengths like in Nd breakup
processes.

In Fig.~\ref{f12}
%In the second group of figures
we compare the full calculations
to the separate decay rates for n-- and p--induced processes.
Note that in this figure
%Fig.~\ref{f12} 
PN means that the proton is nucleon~1 and a neutron is nucleon~2,
which corresponds to $\nu_1 = \frac12$, $\nu_2 = -\frac12$ 
and $\nu_3 = -\frac12$ in the matrix element of Eq.~(\ref{e8.6}).

At $\Theta_{12}=180^o$
the decay rates in the center of the S-curve are essentially given
 by the
p--induced process if a proton neutron pair is registered and
 by the n-- induced process if two neutrons are registered.
Already at $\Theta_{12} = 160^o$ this is no
longer true. The p--induced rate and even more the n--induced rate
is significantly lower than the rate built up by the full
physical process.
It is interesting to see in cases of Fig.~\ref{f12}(b--f) that
the n--induced decay rate for pn detection at 
the upper end of $S$ is practically identical to the decay rate
corresponding to the full physical process.
The reason is that at the upper end of $S$ the energy
$E_1$ (the proton energy) is nearly zero (see Fig.~\ref{f1}), 
and therefore
two neutrons carry essentially all the energy. Without FSI 
such a case can only be generated by a n--induced decay
and qualitatively this picture does not change due to FSI.
In case of p--induced decay its rate for pn and nn detection at the
lower end of $S$ is practically identical with the full decay rate.
The reason is similar as for the n--induced decay. 
At the lower end of $S$ the energy $E_2$ (a neutron energy) is nearly zero.
Therefore a proton--neutron pair has to carry essentially all the energy
and this has to be generated by a p--induced process.
All these enhancements at the lower and upper end of $S$ 
are QFS-like cases.

Interesting are also the FSIP's, especially pronounced at
$\Theta_{12}=0^o$ and $180^o$. At $\Theta_{12}=0^o$ and for
 neutron-neutron detections the proton
has to fly in the opposite direction, therefore the p--induced process
has to be mainly responsible for the peak, as is the case. For
p-n detection, however, both p-- and n--induced decays can contribute
to a FSIP and they do. Apparently the p-- and n--induced decays
have to interfere, since the individual rates do not add up to
the total physical decay.
As a further example we comment on the left strong FSIP at $\Theta_{12}
=180^o$. There are two peaks according to pn and nn detections. 
For nn detection (the neutrons have opposite momenta) this has to be 
necessarily a pn FSIP.
As we see from the figure it receives 
contributions
from n-- and p--induced processes, again coherently. In case of pn detection
(the proton an a neutron have opposite momenta) the special
location on the $S$-curve (high proton energy) requires 
that it is a nn FSIP.
Therefore the very dominant contribution has to come from 
the p--induced process.
This is clearly visible in Fig.~\ref{f12}.

The corresponding curves, when all mesons are included, are qualitatively
the same and only very few examples are displayed in Fig.~\ref{f12.5}.
The heights of the FSIP's have changed, however, significantly.

Finally we consider the question, whether the total n-- and p--induced
decay rates can be separated experimentally. As we already saw this
is not possible in the nd decay channel. There the two processes
interfere coherently
and the individual theoretical rates do not sum up to the total rate.
In the 3N decay channel the total rate is very close to the sum of the
individual rates for the n-- and p--induced processes.
Also in the $ p-\theta$ representation of $ d\Gamma$ (see Eq.~(\ref{e8.5}))
and  displayed in Figs.~\ref{f9}--\ref{f10}
the events from the two different processes appear to be nicely
separated. On the other hand, in the $\Theta_{12} - S$ representation,
which is directly
accessible using two detectors, we saw cases where an interference
was manifest. The total decay rate into 3 nucleons is
\begin{equation}
\Gamma^{\rm n+n+p} \ = \ \int \frac{d^5 \sigma}{d {\hat k}_1 d {\hat k}_2 d S}
\, d {\hat k}_1 d {\hat k}_2 d S \ = \
8 \pi^2 \, \int_0^\pi \, d \Theta_{12} \sin \Theta_{12}
\int_0^{S_{\rm max}(\Theta_{12})} d S \, \sigma (\Theta_{12}, S)
\label{ess}
\end{equation}
We used the fact that
%$\frac{d^5 \sigma}{d {\hat k}_1 d {\hat k}_2 d S}$
${d^5 \sigma} / {d {\hat k}_1 d {\hat k}_2 d S}$
depends only on $\Theta_{12}$ and $S$ and
introduced the length $S_{\rm max}(\Theta_{12})$ of the S-curve depending
on $\Theta_{12}$. First of all we notice immediately that the pure QFS
cases, where a final nucleon momentum is zero, do not contribute,
since for ${\vec k}_3 = 0 $ $\Theta_{12}=180^o$ 
and for ${\vec k}_1 = 0 $ or ${\vec k}_2 = 0 $ the phase space factor
in Eq.~(\ref{e8.6}) is zero. 
Nevertheless an angular configuration
with $\Theta_{12}=180^o$ can and should be measured by itself, since along the
S-curve there will be one point with the exact QFS conditions
and as we saw in Fig.~\ref{f12} there the n-- and p--induced processes can be
cleanly separated. One measures
either a pn or a nn pair and they are generated by p-- and n--induced
decays, respectively.

We now discuss the quantity
\begin{equation}
\gamma (\Theta_{12}) \, = \, 8 \pi^2 \, \sin \Theta_{12}
\int_0^{S_{\rm max}(\Theta_{12})} d S \, \sigma (\Theta_{12}, S)
\label{ess2}
\end{equation}
for np and nn detection, respectively.
This is shown in Figs.~\ref{f13}--\ref{f14}
together with the individual contributions of the p-- and n--induced processes.
We see strong peaks near $\Theta_{12} = 170^o$ for the full processes.
The corresponding values for the p(n)--induced decay are similar
in the peak area for pn(nn) detection, while the n(p)--induced values
are small.

At smaller angles $\Theta_{12}$ the p-- and n--induced values are
similar to each other  in case of pn detections, while for nn detection
the p--induced quantities dominate.
% at the smaller $\Theta_{12}$ values.
A closer inspection reveals that the sum of the p-- and n--induced
values for each $\Theta_{12}$ do not sum up very well to the value
according to the true physical process, but there are differences
up to 10 \% (nn detection) around $\Theta_{12}= 170^o$.
This is a clear signal for interference.

Let us quantify this question. The representation (\ref{e8.5}) of
$ d \Gamma^{n + n +p} / dp d \theta$, which has been displayed
in Figs.~\ref{f4}--\ref{f10} yields the decay rates for the individual
n-- and p--induced processes as well as for the full physical process
(neutron and proton induced)
when integrated over the whole $p$--$\theta$ plane. According to
Figs.~\ref{f9} and~\ref{f10} the n-- and p--induced processes appear
to receive their contributions from well separated areas in the
$p$--$\theta$ plane. Quantitatively, however, this is not true.
Restricting the integration in $\theta$ and $p$ to the region
where the peaks in Figs.~\ref{f9}--\ref{f10} are located  results
in only a certain fraction of the full rates.
Quantitatively, if we fix that fraction to 60~\%, say, for p-- and
n--induced decays, respectively, we find that the regions displayed in
Fig.~\ref{f15} contribute. 
In choosing a certain fraction we always start integration
from the highest values (located in the peaks) 
downwards and stop when the assumed fraction has been reached.
Except for a small domain
($\theta  \approx \pi$ and $p$ large) the regions
for p-- and n--induced decays are clearly separated.
Of course for fractions smaller than 60~\% this will be even more the case.
For fractions larger than 60~\%, however, the regions overlap considerably.
An example for 90~\% is also shown in Fig.~\ref{f15}.
Clearly in such a case the events coming from
for p-- and n--induced decays cannot be separated any more experimentally.

We show the fractional decay rates evaluated over corresponding
increasing regions in Tables~\ref{t8}, \ref{t9}
and in Figs.~\ref{f15}--\ref{f17},
where the results refer to full calculations and include 
all meson exchanges.
There the p-- and n--induced rates are compared to the observed rate
${\Gamma}_{\rm physical}$, produced by the full physical process.
Thus if we require  that the measured value is equal to the p--
or n--induced decay within a few percent, one has to restrict the
integration in the $\theta$--$p$ plane to certain subregions
and the rates to only about 60~\% of the full rate.

Let us map the $\theta$--$p$ values into the
variables $\Theta_{12}$--$S$, which are directly accessible experimentally.
This is shown in Fig.~\ref{f17} for the 60~\% and 90~\%.
That picture refers to the detection
of a neutron (particle~1) -- proton (particle~2) pair.
Fig.~\ref{f17} tells that n--induced decay can be found
under all $\Theta_{12}$ angles for small $S$-values. ($E_2$ is then small.)
The p--induced decay on the other hand is to be found
for all the $\Theta_{12}$-angles
around maximal $S$-values and in the region
$160^o \le \Theta_{12} \le 180^o $ for medium-large $S$-values.
(For small neutron energies.)
A correspondingly modified figure could be shown
if two neutrons would be detected.
If on the other hand 90~\% of the corresponding rates should be detected then
the two detectors would receive events from both processes under the same
angle $\Theta_{12}$ and for the same energies in a large portion
of the phase space. Thus experimentally it is not possible
to separate those processes.

We have to conclude that despite the fact that $\Gamma^{n + n +p}$
is rather close to the sum of $\Gamma^{n + n +p}_{(p)}$
and $\Gamma^{n + n +p}_{(n)}$,
the latter values cannot be determined experimentally,
only fractions, the smaller, the cleaner.

\section{Summary and Conclusions}
\label{secVI}

For the first time the nonmesonic hypertriton decay has been calculated
based on rigorous solutions of 3-body Faddeev equations for the hypertriton
and the 3N scattering states of the final 3 nucleons. Realistic NN and
hyperon-nucleon interactions have been used. In the meson exchange process
the pion exchange is dominant, but the other included mesons $\eta$, K,
$\omega$, $\rho$ and K$^*$ provide also significant
contributions of various signs and magnitudes.
The final state interaction turned out to be very important
and reduces the rates for PWIAS by
about a factor of $2$. 
The total nonmesonic decay rate turns out to be
$0.55 \%$ of the free $\Lambda$ decay rate and is smaller by an order
of magnitude than a previous estimation[4] which used the pion-exchange
model, a much more simplified hypertriton wavefunction and no FSI.
While the p-- and n--induced decays add up
manifestly in a coherent manner in the nd decay process, the nnp decay
rate is rather well given as the sum of the n-- and p--induced decay rates.
Nevertheless these individual decay rates cannot be measured separately.
Only fractions thereof can be obtained, when the contributions arise from
non-overlapping regions in phase space. This subject has been thoroughly
discussed in section~\ref{secV}. Detailed information has been given
regarding the separation of
p-- and n--induced differential decay rates experimentally.

Since the decay rates depend sensitively on the number and type of mesons
exchanged, it will be an interesting testground for the dynamics of these
meson exchanges, which are driven by weak and a strong vertices. At the
same time the decay rates probe the hypertriton wavefunction and the reaction
mechanism of the three outgoing nucleons through their strong final
state interaction.
The latter one is especially manifest in the strong FSIP's, where two
nucleons leave with equal momenta.

The evaluation of the pionic decay into the various bound states and
continuum channels can be performed in a similar manner and is planned.

\acknowledgements

This work was supported by the
Polish Committee for Scientific Research under Grant No.~PB~1031,
the Science and Technology Cooperation Germany-Poland
under Grant No.~XO81.91, the DGICYT Grant No. PB92-0761 (Spain),
the Generalitat de Catalunya Grant No.  GRQ94-1022 and the US-DOE
Grant No. DE-FG02-95-ER40907. AP acknowledges support from a doctoral
fellowship of the Ministerio de Educaci\'on y Ciencia (Spain).
The numerical calculations have been performed on the Cray Y-MP of the
H\"ochstleistungsrechenzentrum in J\"ulich, Germany.

\newpage

\begin{figure}
\caption{Locus for kinematically allowed events
in the $E_1$--$E_2$ plane and $\Theta_{12} = 180^o$
together with our definition of the choice for the arclength $S = 0$.
From that point on $S$ is evaluated for each point on the locus in the counterclockwise
sense.}
\label{f1}
\end{figure}

\begin{figure}
\caption{The $\pi$--induced $\Lambda N \rightarrow N N$  exchange process
of Eqs.~(\protect\ref{e20}--\protect\ref{e21}).}
\label{f1.5}
\end{figure}

\begin{figure}
\caption{Individual meson contributions to the
isoscalar (a) and isovector (b) central spin independent regularized 
potential $r^2 V^{\mu}_{C}(r)$. The potential obtained by
adding all meson contributions for the isoscalar (c) and isovector (d)
case.}
\label{fig:c}
\end{figure}

\begin{figure}
\caption{Individual meson contributions to the 
isoscalar (a) and isovector (b) central spin dependent regularized 
potential $r^2 V^{\mu}_{SS}(r)$. The potential obtained by 
adding all meson contributions for the isoscalar (c) and isovector (d)
case.}
\label{fig:ss}
\end{figure}

\begin{figure}
\caption{Individual meson contributions to the
isoscalar (a) and isovector (b) tensor regularized 
potential $r^2 V^{\mu}_{T}(r)$. The potential obtained by
adding all meson contributions for the isoscalar (c) and isovector (d)
case.}
\label{fig:t}
\end{figure}

\begin{figure}
\caption{Individual meson contributions to the
isoscalar (a) and isovector (b) parity violating regularized 
potential $r^2 V^{\mu}_{PV}(r)$. The potential obtained by
adding all meson contributions  for the isoscalar (c) and isovector (d)
case.}
\label{fig:pv}
\end{figure}

\begin{figure}
\caption{The location of the 3 peaks corresponding
to ${\vec k}_i = 0 $ (i=1,2,3) and the FSI peaks
in the $\theta$--$p$ plane (see text).}
\label{f2}
\end{figure}

\begin{figure}
\caption{The differential decay rate $ d \Gamma^{n + n +p} / dp d \theta$
for PWIAS and exchange of all mesons.}
\label{f4}
\end{figure}

%\begin{figure}
%\caption{The same as in Fig.~\protect\ref{f3} for exchange of all mesons.}
%\label{f4}
%\end{figure}

\begin{figure}
\caption{The n--induced differential decay rate
$ d \Gamma^{n + n + p}_{(n)} / dp d \theta$ for PWIAS
and $\pi$-exchange only. Nucleons 1, 2 and 3 are chosen to be
a neutron, a proton and a neutron.}
\label{f5}
\end{figure}

\begin{figure}
\caption{The same as in Fig.~\protect\ref{f5} for p--induced decay.}
\label{f6}
\end{figure}

\begin{figure}
\caption{The differential decay rate $ d \Gamma^{n + n +p} / dp d \theta$
with full inclusion of the final state interaction
and exchange of all mesons.}
\label{f8}
\end{figure}

%\begin{figure}
%\caption{The same as in Fig.~\protect\ref{f7} for the exchange of all mesons.}
%\label{f8}
%\end{figure}

\begin{figure}
\caption{The n--induced differential decay rate
$ d \Gamma^{n + n +p}_{(n)} / dp d \theta$ with full inclusion of
the final state interaction and $\pi$--exchange only.}
\label{f9}
\end{figure}

\begin{figure}
\caption{The same as in Fig.~\protect\ref{f9} for p--induced decay.}
\label{f10}
\end{figure}

\begin{figure}
\caption{The differential decay rate
$ d \Gamma^{n + n + p} / d \Theta_{12} dS $ for various
angles $\Theta_{12}$ and $\pi$--exchange only.
The two detected nucleons are either a p(particle~1)n(particle~2)
pair or two neutrons.
PWIAS is compared to the treatment including the full final state interaction.}
\label{f11}
\end{figure}

\begin{figure}
\caption{The n--and p--induced differential decay rates
$ d \Gamma^{n + n +p}_{(n),(p)} / d \Theta_{12} dS $ in comparison
to the physical rate for various
angles $\Theta_{12}$ and $\pi$--exchange only.
The two detected nucleons are either a p(particle~1)n(particle~2)
pair or two neutrons.
The final state interaction is fully included.}
\label{f12}
\end{figure}

\begin{figure}
\caption{The same as in Fig.~\protect\ref{f12} for the exchange of all mesons.}
\label{f12.5}
\end{figure}

\begin{figure}
\caption{Differential decay rates integrated over the S-curve as a
function of $\Theta_{12}$. The p-- and n--induced cases are compared
to the physical process. The final state interaction is fully
included and all mesons are exchanged. The curves belong to the case
that a proton and a neutron are detected.}
\label{f13}
\end{figure}

\begin{figure}
\caption{The same as in Fig.~\protect\ref{f13} for the case that two neutrons
are detected.}
\label{f14}
\end{figure}

\begin{figure}
\caption{The separate regions in the $\theta$--$p$ plane contributing to
(a) 60~\% and (b) 90 \% of the rates of n-- and p--induced decays. 
Note the strong overlap of the different processes in phase space in case (b).}
\label{f15}
\end{figure}

%\begin{figure}
%\caption{The same as in Fig.~\protect\ref{f15} for the case of 90~\%.
%Note the strong overlap of the different processes in phase space.}
%\label{f16}
%\end{figure}

\begin{figure}
\caption{The separate regions in the $\Theta_{12}$--$S$ plane contributing to
(a) 60~\% and (b) 90~\% of the rates of n-- and p--induced decays.
Note the strong overlap of the different processes in phase space in case (b).
(Particle~1 is a neutron and particle~2 is a proton.)}
\label{f17}
\end{figure}

%\begin{figure}
%\caption{The same as in Fig.~\protect\ref{f17} for the case of 90~\%.
%Note the strong overlap of the different processes in phase space.}
%\label{f18}
%\end{figure}

\newpage

\begin{table}
\centering
\caption{Constants appearing in the weak transition potential for
different mesons (in units of $G_F {m_\pi}^2$). The strong and weak
coupling constants have been taken from Ref.\protect\cite{7}}
\vskip 0.2 in
\begin{tabular}{lcccc}
 $\mu$  & $K^\mu_{C}$ & $K^\mu_{SS}$ & $K^\mu_{T}$ & $K^\mu_{PV}$ \\
\hline\hline
&   &    &       &         \\
$\pi$ & 0 & $\dfrac{B_\pi}{2 \overline M}
\dfrac{g_{\rm {\sst NN} \pi}}{2M_N}$ &
 $\dfrac{B_\pi}{2 \overline M}
\dfrac{g_{\rm {\sst NN} \pi}}{2M_N}$ &
$A_\pi  \dfrac{g_{\rm {\sst NN} \pi}}{2M_N}$ \\
&   &    &       &         \\
$\eta$ & $0$ & $\dfrac{B_\eta}{2 \overline M}
\dfrac{g_{\rm {\sst NN} \eta}}{2M_N}$ &
$\dfrac{B_\eta}{2 \overline M}
\dfrac{g_{\rm {\sst NN} \eta}}{2M_N}$ &
$A_\eta  \dfrac{g_{\rm {\sst NN} \eta}}{2M_N}$ \\
&    &     &       &         \\
$K$ & $0$ & $\dfrac{1}{2M_N} \dfrac{g_{\rm {\sst \Lambda N K}}}
{2 \overline M}$ &
$\dfrac{1}{2 M_N}
\dfrac{g_{\rm \sst{\Lambda N K}}}{2 \overline M}$
& $\dfrac{g_{\rm \sst{\Lambda N K}}}{2M_N}$
\\
&    &     &       &         \\
$\rho$ & $g^{\rm {\sst V}}_{\rm {\sst NN} \rho} \alpha_\rho$ &
$2\dfrac{\alpha_\rho + \beta_\rho}{2 \overline M}
\dfrac{g^{\rm {\sst V}}_{\rm {\sst NN} \rho} +
g^{\rm {\sst T}}_{\rm {\sst NN} \rho}} {2M_N}$ &
$ - \dfrac{\alpha_\rho + \beta_\rho}{2 \overline M}
\dfrac{ g^{\rm {\sst V}}_{\rm {\sst NN} \rho} +
g^{\rm {\sst T}}_{\rm {\sst NN} \rho}} {2M_N}$ &
$ - \varepsilon_\rho
\dfrac{ g^{\rm {\sst V}}_{\rm {\sst NN} \rho} +
g^{\rm {\sst T}}_{\rm {\sst NN} \rho}} {2M_N}$ \\
&    &     &       &         \\
$\omega$ & $g^{\rm {\sst V}}_{\rm {\sst NN} \omega} \alpha_\omega$ &
$2\dfrac{\alpha_\omega + \beta_\omega}{2 \overline M}
\dfrac{g^{\rm {\sst V}}_{\rm {\sst NN} \omega} +
g^{\rm {\sst T}}_{\rm {\sst NN} \omega}} {2M_N} $ &
$ - \dfrac{\alpha_\omega + \beta_\omega}{2 \overline M}$
$\dfrac{ g^{\rm {\sst V}}_{\rm {\sst NN} \omega} +
g^{\rm {\sst T}}_{\rm {\sst NN} \omega}} {2M_N} $ &
$ - \varepsilon_\omega
\dfrac{ g^{\rm {\sst V}}_{\rm {\sst NN} \omega} +
g^{\rm {\sst T}}_{\rm {\sst NN} \omega}} {2M_N}$ \\
&    &     &       &         \\
$K^*$ & $g^{\rm {\sst V}}_{\rm {\sst \Lambda N K^*}} $ &
$2 \dfrac{1}{2M_N} \dfrac{g^{\rm \sst{V}}_{\rm {\sst \Lambda N K^*}} +
g^{\rm \sst{T}}_{\rm {\sst \Lambda N K^*}}} {2 \overline M}$  &
$ - \dfrac{1}{2M_N}
\dfrac{ g^{\rm {\sst V}}_{\rm {\sst \Lambda N K^*}} +
g^{\rm \sst{T}}_{\rm {\sst \Lambda N K^*}}} {2 \overline M}$ &
$- \dfrac{ g^{\rm \sst{V}}_{\rm {\sst \Lambda N K^*}} +
g^{\rm \sst{T}}_{\rm {\sst \Lambda N K^*}}} {2M_N}$ \\
&    &     &       &         \\
\end{tabular}
\label{tab:k}
\end{table}

\begin{table}
\centering
\caption{Strong (Nijmegen) and weak coupling constants
and cutoff parameters for the different mesons.
The weak couplings are in units of $G_F {m_\pi}^2 = 2.21 \times 10^{-7} $.
}
\vskip 0.2 in
\begin{tabular}{|c|l|l|l|l|}
Meson & \multicolumn{1}{c|}{Strong c.c.} & \multicolumn{2}{c|}{Weak c.c.} &
$\Lambda_i$ \\
      &     & \multicolumn{1}{c}{PC} & \multicolumn{1}{c|}{PV} &
\mbox{(GeV)} \\ \hline
 $\pi$ & $g_{\rm {\scriptscriptstyle NN} \pi}$ = 13.3 & $B_\pi$=$-$7.15 &
$A_\pi$=1.05 & 1.30  \\
    & $g_{\rm {\scriptscriptstyle \Lambda \Sigma} \pi}$ = 12.0 &  &  &  \\
\hline
 $\eta$ & $g_{\rm {\scriptscriptstyle NN} \eta}$ = 6.40 & $B_\eta$=$-$14.3 &
 $A_\eta$=1.80 & 1.30 \\
   & $g_{\rm {\scriptscriptstyle \Lambda \Lambda} \eta}= -6.56$ &  &  &  \\
\hline
K & $g_{\rm {\scriptscriptstyle \Lambda NK}}$ = $-$14.1 &
$C_{\rm {\scriptscriptstyle K}}^{\rm {\scriptscriptstyle PC}}$=$-$18.9
& $C_{\rm {\scriptscriptstyle K}}^{\rm {\scriptscriptstyle PV}}$=0.76 & 1.20 \\
    & $g_{\rm {\scriptscriptstyle N \Sigma K}}$ = 4.28 & $D_{\rm
 {\scriptscriptstyle K}}^
{\rm {\scriptscriptstyle PC}}$=6.63 & $D_{\rm {\scriptscriptstyle K}}^{\rm
 {\scriptscriptstyle PV}}$=2.09  & \\
\hline
$\rho$ & $g_{\rm {\scriptscriptstyle NN} \rho}^{\rm {\scriptscriptstyle V}}$ =
 3.16 &
$\alpha_\rho$=$-$3.50 & $\epsilon_\rho$=1.09  & 1.40 \\
       & $g_{\rm {\scriptscriptstyle NN} \rho}^{\rm {\scriptscriptstyle T}}$ =
 13.3 &
$\beta_\rho$=$-$6.11   &  &  \\
   & $g_{\rm {\scriptscriptstyle \Lambda \Sigma} \rho}^{\rm {\scriptscriptstyle
 V}}= 0$ &
   &  & \\
   & $g_{\rm {\scriptscriptstyle \Lambda \Sigma} \rho}^{\rm {\scriptscriptstyle
 T}}=
   11.2$ &  &  &  \\
\hline
$\omega$ & $g_{\rm {\scriptscriptstyle NN} \omega}^{\rm {\scriptscriptstyle V}}$
 = 10.5
& $\alpha_\omega$=$-$3.69  & $\epsilon_\omega$= $-$1.33
& 1.50 \\
      & $g_{\rm {\scriptscriptstyle NN} \omega}^{\rm {\scriptscriptstyle T}}$ =
 3.22 &
$\beta_\omega$=$-$8.04  &  & \\
   & $g_{\rm {\scriptscriptstyle \Lambda \Lambda} \omega}^{\rm
 {\scriptscriptstyle V}}=
   7.11$ & &  &  \\
   & $g_{\rm {\scriptscriptstyle \Lambda \Lambda} \omega}^{\rm
 {\scriptscriptstyle T}}=
   -4.04$ & &  &  \\
\hline
K$^*$ & $g_{\rm {\scriptscriptstyle \Lambda N K^*}}^{\rm
{\scriptscriptstyle V}}$ =
$-$5.47 & $C^{\rm {\scriptscriptstyle PC,V}}_{\rm {\scriptscriptstyle
K^*}}$=$-$3.61 &
$C^{\rm {\scriptscriptstyle PV}}_{\rm {\scriptscriptstyle K^*}}$=$-$4.48 & 2.20
 \\
     & $g_{\rm {\scriptscriptstyle \Lambda N K^*}}^{\rm {\scriptscriptstyle T}}$
 =
$-$11.9 &
$C^{\rm {\scriptscriptstyle PC,T}}_{\rm {\scriptscriptstyle K^*}}$=$-$17.9
&   & \\
   &$g_{\rm {\scriptscriptstyle N \Sigma K^*}}^{\rm {\scriptscriptstyle
 V}}=-3.16$ &
   $D^{\rm {\scriptscriptstyle PC,V}}_{\rm {\scriptscriptstyle K^*}}$=$-$4.89 &
   $D^{\rm {\scriptscriptstyle PV}}_{\rm {\scriptscriptstyle K^*}}$=0.60 & \\
   &$g_{\rm {\scriptscriptstyle N \Sigma K^*}}^{\rm {\scriptscriptstyle
 T}}=6.00$ &
   $D^{\rm {\scriptscriptstyle PC,T}}_{\rm {\scriptscriptstyle K^*}}$=9.30 &  &
 \\
\end{tabular}
\label{tab:const}
\end{table}

\begin{table}
\caption{Decay rates in units of $s^{-1}$ for individual meson exchanges 
and for partially summed up exchanges.}
\begin{center}
\begin{tabular} {|c|c||l|c|} 
\multicolumn{2}{|c||}{${\Gamma}^{\rm meson} $} &
\multicolumn{2}{c|}{${\Gamma}^{\rm \, partially \ summed} $} \\[4pt] \hline
$\pi$ & 0.2412 $\times 10^8$  & $\pi$ & 0.2412 $\times 10^8 $  \\[4pt] \hline
$\eta $ & 0.4826 $\times 10^6$ & $\pi \, + \, \eta $ & 0.2299 $\times 10^8 $ \\[4pt] \hline
$K $ & 0.5422 $\times 10^7 $ & $\pi \, + \, \eta \, + \, K $ & 0.9267 $\times 10^7$   \\[4pt] \hline
$\rho $ & 0.7647 $\times 10^6 $ & $\pi \, + \, \eta \, + \, K \, + \, \rho $ & 0.7502 $\times 10^8 $  \\[4pt] \hline
$\omega $ & 0.4372 $\times 10^7 $ & $\pi \, + \, \eta \, + \, K \, + \, \rho \,
+ \, \omega $ & 0.1752 $\times 10^8$  \\[4pt] \hline
$K^* $ & 0.5569 $\times 10^7  $ & $\pi \, + \, \eta \, + \, K \, + \, \rho \, + \, \omega \,
+ \, K^*$ & 0.2126 $\times 10^8 $ \\[4pt]
\end{tabular}
\end{center}
\label{t1}
\end{table}

\begin{table}
\caption{Selected decay rates in units of $s^{-1}$ for $\pi$--exchange only
and for exchange of all mesons.}
\begin{center}
\begin{tabular} {|c|c||c|} 
\multicolumn{1}{|c|}{} &
\multicolumn{1}{c||}{$\pi$--exchange only\ \ \ } &
\multicolumn{1}{c|}{Exchange of all mesons\ \ \ } \\[4pt] \hline
${\Gamma}^{\rm \, n+d}_{\rm PWIAS}   $ & 0.59 $ \times 10^7 $  & 0.47 $ \times 10^7 $ \\[4pt] \hline
${\Gamma}^{\rm \, n+d}               $ & 0.15 $ \times 10^7 $  & 0.22 $ \times 10^7 $ \\[4pt] \hline
${\Gamma}^{\rm \, n+n+p}_{\rm PWIAS}  $ & 0.46 $ \times 10^8 $ & 0.36 $ \times 10^8 $ \\[4pt] \hline
${\Gamma}^{\rm \, n+n+p}              $ & 0.23 $ \times 10^8 $ & 0.19 $ \times 10^8 $ \\[4pt] \hline
${\Gamma}^{\rm \, n+d}_{\rm PC}      $ & 0.88 $ \times 10^6 $ & 0.22 $ \times 10^7 $  \\[4pt] \hline
${\Gamma}^{\rm \, n+d}_{\rm PV}      $ & 0.59 $ \times 10^6 $ & 0.22 $ \times 10^5 $  \\[4pt] \hline
${\Gamma}^{\rm \, n+p+p}_{\rm PC}     $ & 0.13 $ \times 10^8 $  & 0.12 $ \times 10^8 $ \\[4pt] \hline
${\Gamma}^{\rm \, n+p+p}_{\rm PV}     $ & 0.92 $ \times 10^7 $ & 0.73 $ \times 10^7 $  \\[4pt] 
\end{tabular}
\end{center}
\label{t2it3}
\end{table}

\begin{table}
\caption{Proton-- and neutron-- induced decay rates 
in units of $s^{-1}$ for $\pi$--exchange only
and for the exchange of all mesons in comparison to the total nd and nnp 
rates.}
\begin{center}
\begin{tabular} {|c|c||c|}
\multicolumn{1}{|c|}{} &
\multicolumn{1}{c||}{$\pi$--exchange only\ \ \ } &
\multicolumn{1}{c|}{Exchange of all mesons\ \ \ } \\[4pt] \hline
${\Gamma}^{\rm \, n+d}_{\rm (n)}     $ & 0.20 $ \times 10^6 $ & 0.64 $ \times 10^6 $ \\[4pt] \hline
${\Gamma}^{\rm \, n+d}_{\rm (p)}     $ & 0.78 $ \times 10^6 $ & 0.70 $ \times 10^6 $ \\[4pt] \hline
${\Gamma}^{\rm \, n+d}               $ & 0.15 $ \times 10^7 $ & 0.22 $ \times 10^7 $ \\[4pt] \hline
${\Gamma}^{\rm \, n+n+p}_{\rm (n)}    $ & 0.59 $ \times 10^7 $ & 0.57 $ \times 10^7 $ \\[4pt] \hline
${\Gamma}^{\rm \, n+n+p}_{\rm (p)}    $ & 0.18 $ \times 10^8 $ & 0.13 $ \times 10^8 $ \\[4pt] \hline
${\Gamma}^{\rm \, n+n+p}              $ & 0.23 $ \times 10^8 $ & 0.19 $ \times 10^8 $ \\[4pt]
\end{tabular}
\end{center}
\label{t4it5}
\end{table}

\begin{table}
\caption{Frac\-tio\-nal pro\-ton-- and neu\-tron--in\-du\-ced de\-cay ra\-tes 
in units of $s^{-1}$  
$ \Gamma_{(p),(n)} = \int d \Gamma^{n + n +p}_{(p),(n)} $ integrated  over 
subdomains in the $\theta$--$p$ plane, where proton induced decay dominates.
They are compared to the corresponding fractional physical rate.}
\begin{center}
\begin{tabular} {|c|cccc|} 
  area (P)  &  ${\Gamma}_{\rm (p)}$  
&  ${\Gamma}_{\rm (n)}$   
&  ${\Gamma}_{\rm physical}$  
& ${\Gamma}_{\rm (p)} / {\Gamma}_{\rm physical}$ \\[4pt] \hline
 60 \%      & 0.80 $ \times 10^7 $ &  0.20 $ \times 10^6$  &  0.82 $ \times  10^7 $      & 98 \% \\[4pt] \hline
 70 \%      & 0.94 $ \times 10^7 $ &  0.48 $ \times 10^6$  &  0.98 $ \times  10^7 $      & 96 \% \\[4pt] \hline
 80 \%      & 0.11 $ \times 10^8 $ &  0.99 $ \times 10^6$  &  0.12 $ \times  10^7 $      & 92 \% \\[4pt] \hline
 90 \%      & 0.12 $ \times 10^8 $ &  0.23 $ \times 10^7$  &  0.14 $ \times  10^8 $      & 86 \% \\[4pt]
\end{tabular}
\end{center}
\label{t8}
\end{table}

\begin{table}
\caption{Fractional proton-- and neutron--induced decay rates 
$ \Gamma_{(p),(n)} = \int d \Gamma^{n + n +p}_{(p),(n)} $ in units of $s^{-1}$ integrated  over 
subdomains in the $\theta$--$p$ plane, where neutron induced decay dominates.
They are compared to the corresponding fractional physical rate.}
\begin{center}
\begin{tabular} {|c|cccc|} 
  area (N)  &  ${\Gamma}_{\rm (p)}$  &  ${\Gamma}_{\rm (n)}$   &  ${\Gamma}_{\rm physical}$  & ${\Gamma}_{\rm (n)} / {\Gamma}_{\rm physical}$ \\[4pt] \hline
 60 \%      & 0.34 $ \times 10^6 $ &  0.34 $ \times 10^7$  &  0.35 $ \times  10^7 $      & 97 \% \\[4pt] \hline
 70 \%      & 0.11 $ \times 10^7 $ &  0.40 $ \times 10^7$  &  0.47 $ \times  10^7 $      & 85 \% \\[4pt] \hline
 80 \%      & 0.29 $ \times 10^7 $ &  0.45 $ \times 10^7$  &  0.72 $ \times  10^7 $      & 63 \% \\[4pt] \hline
 90 \%      & 0.50 $ \times 10^7 $ &  0.51 $ \times 10^7$  &  0.10 $ \times  10^8 $      & 51 \% \\[4pt]
\end{tabular}
\end{center}
\label{t9}
\end{table}


\begin{references}
\bibitem{1} K.~Miyagawa, H.~Kamada, W.~Gl\"ockle, V.~G.~J.~Stoks,
Phys. Rev. {\bf C 51} (1995) 2905.
\bibitem{2} P.~M.~M.~Maessen, Th.~A.~Rijken, J.~J.~de Swart,
Phys. Rev. {\bf C 40} (1989) 2226.
\bibitem{3} G. Keyes, et al, Nucl. Phys. {\bf B67} (1973) 269.
\bibitem{4} C.~Bennhold, A.~Ramos, D.~A.~Aruliah, and U.~Oelfke,
Phys. Rev. {\bf C 45} (1992) 947.
\bibitem{5} J.~Cohen, Progr. Part. Nucl. Phys. {\bf 25} (1990) 139.
\bibitem{6} J.F. Donoghue et al., Phys. Rep. {\bf 131} (1986) 319;
L. de la Torre, Ph.D. Thesis, Univ. of Massachusetts (1982); J. Dubach
et al, Annals Phys. (in print).
\bibitem{7} A.~Parre\~no, A.~Ramos, C.~Bennhold, submitted to Phys. Rev. C.
\bibitem{fldec} Review of Particle Properties, Phys. Rev. {\bf D 54} (1996) 619
and references therein.
\bibitem{8} W.~Gl\"ockle, H.~Wita{\l}a, D.~H\"uber, H.~Kamada, J.~Golak,
Phys. Report {\bf 274} (1996) 107.
\bibitem{9} J.~Golak,  H.~Kamada,  H.~Wita{\l}a, W.~Gl\"ockle, S.~Ishikawa,
Phys. Rev. {\bf C 51} (1995) 1638.
\bibitem{10} W.~Gl\"ockle, {\it The Quantum Mechanical Few-Body Problem\/}
(Springer, Berlin-Tokyo, 1983).
\bibitem{11} R.~Machleidt, K.~Holinde, Ch.~Elster,
Phys. Rep. {\bf 149} (1987) 1.
\bibitem{12} V.~G.~J.~Stoks, R.~A.~M.~Klomp, C.~P.~F.~Terhessen, J.~J.~de Swart,
Phys. Rev. {\bf C 49} (1994) 2950.
\end{references}
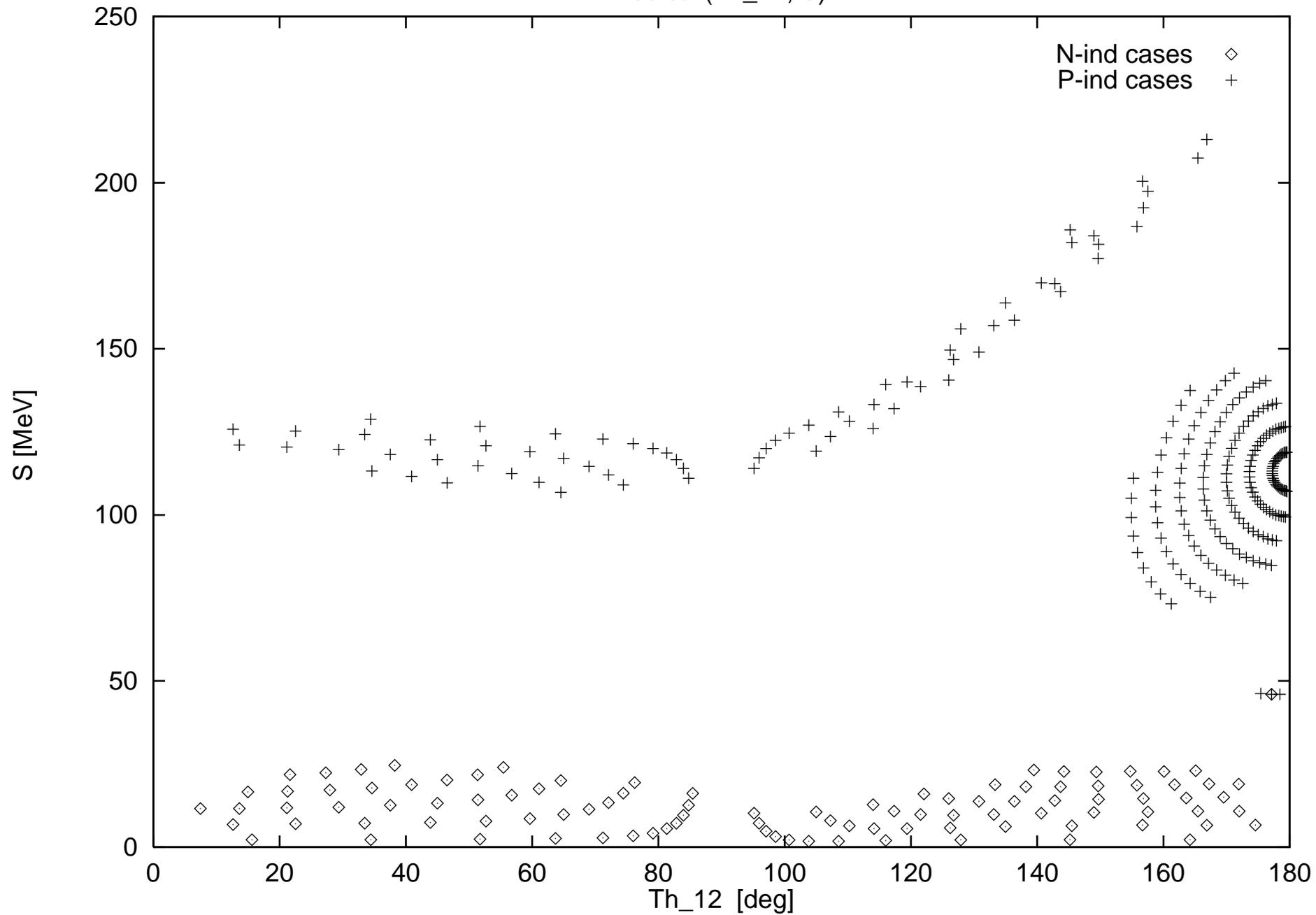
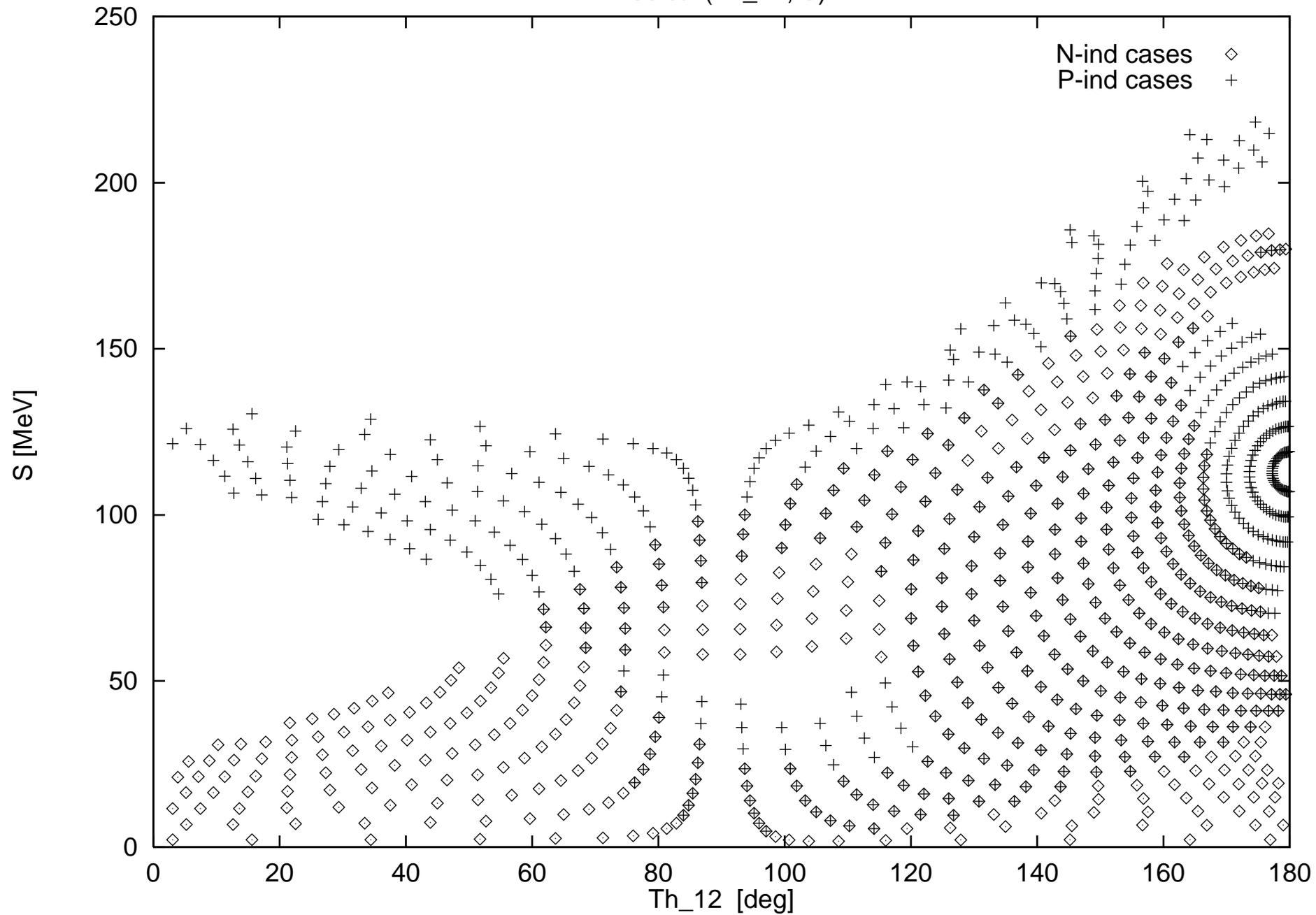
\end{document}